\documentclass[preprint,12pt]{elsarticle}

\usepackage{color}
\usepackage{url}
\usepackage{multirow}

\usepackage{amssymb}

\journal{X-ray Optics and Instrumentation}

\newcommand{\degr} {$^\circ$}
\newcommand{\ttwopi}{$t_{2\pi}$}
\newcommand{\tfourpi}{$t_{4\pi}$}
\newcommand{\tsixpi}{$t_{6\pi}$}
\newcommand{\arcsec}{$''$}
\newcommand{\muas}{$\mu''$}
\newcommand{\micron}{$\mu$m}

\def \Msun{M$_\odot$}

\newcommand{\etal}{{\it et al.}}
\newcommand{\arzoumanian}{2004SPIE.5488..623A}
\newcommand{\baez}{1952JOSA...42..756B}
\newcommand{\bennett}{1976ApOpt..15..542B}
\newcommand{\braiga}{2004SPIE.5488..601B}
\newcommand{\braigb}{2006ExA....21..101B}
\newcommand{\braigc}{2007ApOpt..46.2586B}
\newcommand{\braigd}{2007SPIE.6688E...7B}
\newcommand{\braige}{2010ExA....27..131B}
\newcommand{\brauninger}{1971SoPh...20...81B}
\newcommand{\burger}{1972SoPh...24..395B}

\newcommand{\cao}{2000JOSAA..17..447C}

\newcommand{\cash}{2005AdSpR..35..122C}
\newcommand{\chao}{2008SPIE.6883E...9C}

\newcommand{\desai}{1993SPIE.1948...75D}
\newcommand{\dewey}{1996SPIE.2805..224D}
\newcommand{\difabrizio}{1999Natur.401..895D}
\newcommand{\dijkstra}{1971IAUS...41..207D}

\newcommand{\doeleman}{2008Natur.455...78D}
\newcommand{\elwert}{1968IAUS...35..439E}
\newcommand{\elwertfeitzinger}{1970Optik..31..600E}
\newcommand{\erosita}{2009SPIE.7435E...1M}

\newcommand{\frontera}{frontera}

\newcommand{\gaia}{2007SPIE.6690E...8L}
\newcommand{\gebhardt}{2009ApJ...700.1690G}
\newcommand{\gendreaua}{2003SPIE.4852..685G}

\newcommand{\giaconni}{1965ApJ...142.1274G}
\newcommand{\gillessen}{2009ApJ...692.1075G}
\newcommand{\gimenez}{2006OExpr..1411958G}
\newcommand{\gorensteina}{2003SPIE.4851..599G}
\newcommand{\gorensteinb}{2004SPIE.5168..411G}
\newcommand{\gorensteinc}{2005SPIE.5900..369G}
\newcommand{\gorensteind}{2008SPIE.7011E..23G}
\newcommand{\hyde}{2002SPIE.4849...28H}
\newcommand{\kelley}{2009AIPC.1185..757K}
\newcommand{\kipp}{2001Natur.414..184K}
\newcommand{\kirz}{1974JOSA...64..301K}
\newcommand{\koechlin}{2009Ap&SS.320..225K}
\newcommand{\krichbaumb}{2006evn..confE...2K}
\newcommand{\krichbaum}{2006JPhCS..54..328K}
\newcommand{\krizmanica}{2005ExA....20..299K}
\newcommand{\krizmanicb}{2005ExA....20..497K}

\newcommand{\liq}{2008arXiv0811.3201L}
\newcommand{\lengeler}{2005JPhD...38A.218L}
\newcommand{\mast}{2007RvGeo..45.2004F}
\newcommand{\maximref}{2003SPIE.4852..196C}

\newcommand{\mertz}{1965trop.book.....M}
\newcommand{\michette}{2001SPIE.4145..303M}
\newcommand{\middelberg}{2008RPPh...71f6901M}
\newcommand{\miyamoto}{1961JOSA...51...17M}
\newcommand{\murata}{2007IAUS..242..517M}
\newcommand{\myers}{1951AmJPh..19..359M}
\newcommand{\niemann}{1974PhDT.......154N}
\newcommand{\rayleigh}{rayleigh}
\newcommand{\simpsonmichette}{1984AcOpt..31..403S}
\newcommand{\skinnerao}{2004ApOpt..43.4845S}
\newcommand{\skinnera}{2001A&A...375..691S}
\newcommand{\skinnerb}{2002A&A...383..352S}

\newcommand{\skinnern}{2008SPIE.7011E..22S}
\newcommand{\skinnerk}{2009SPIE.7437E..17S}
\newcommand{\skinnerl}{2009astro2010T..20S}
\newcommand{\skinnerm}{2009ExA....27...61S}
\newcommand{\sochackia}{1992ApOpt..31.5326S}

\newcommand{\soret}{1875AnP...232...99S}

\newcommand{\stahle}{1999PhT....52h..32S}
\newcommand{\stigliani}{1967JOSA...57..610S}
\newcommand{\stonegeorge}{1988ApOpt..27.2960S}
\newcommand{\vaiana}{1977SSI.....3...19V}
\newcommand{\wang}{2003Natur.424...50W}
\newcommand{\white}{2000Natur.407..146W}
\newcommand{\wilson}{1973OptCo...8..384W}
\newcommand{\wunderer}{2006SPIE.6266E..66W}
\newcommand{\xeus}{2006SPIE.6266E..50P}
\newcommand{\yang}{1993NIMPA.328..578Y}
\newcommand{\young}{1972JOSA...62..972Y}

\begin{document}

\begin{frontmatter}



\title{Diffractive X-ray  Telescopes}

\ead{skinner@milkyway.gsfc.nasa.gov}
\author{Gerald K. Skinner}

\address{CRESST  \&  NASA-GSFC, Greenbelt. MD 20771, USA; \\ and \\
 Univ. Md. Department of Astronomy, College Park, MD 20742, USA}

\begin{abstract}

Diffractive X-ray telescopes using zone plates, phase Fresnel lenses, or related optical elements have the potential to provide astronomers with true imaging capability with resolution several orders of magnitude better than available in any other waveband. Lenses that would be relatively easy to fabricate could have an angular resolution of the order of micro-arc-seconds or even better, that would allow, for example, imaging of the distorted space-time in the immediate vicinity of the super-massive black holes in the center of active galaxies What then is precluding their immediate adoption? Extremely  long focal lengths, very limited bandwidth, and difficulty stabilizing the image are the main problems. The history, and status of the development of such lenses is reviewed here and the prospects for managing the challenges that they present are discussed.

\end{abstract}


\end{frontmatter}

\section{INTRODUCTION}
\label{sec:intro} 
Diffractive optics, in the form of zone plates and various forms of Fresnel lenses and kineforms, already play a major role in the manipulation of X-ray beams at synchrotron facilities and in X-ray microscopy. Diffractive X-ray telescopes, in contrast, exist almost entirely as concepts on paper and as proposals and suggestions, though as will be seen demonstrations of scaled systems have been made. Because of atmospheric absorption, their potential application is almost certainly limited to  astronomy (and specifically to astronomy from space), perhaps including solar and planetary studies. However, for certain objectives within that field they present the prospect of enormous advances over current instrumentation, which relies largely on grazing incidence reflective optics (reviewed elsewhere in this series \cite{reflectingreview, reflectingreview2, reflectingreview3}).  The most notable prospect that diffractive telescopes offer is that of superb angular resolution, with improvements of perhaps six orders of magnitude on the current state of the art. But even neglecting benefits from the imaging properties, their capability of concentrating the flux received over a large effective area onto a small, and hence low background, detector may also offer unique advantages in some circumstances.    

This review will consider the various concepts that have been proposed and discuss  the current state of development of the technologies necessary to turn the ideas into a real system. For simplicity the term X-rays will often be used to apply to both X-rays and gamma-rays, there being no clear distinction or borderline between the two.  The review will be limited to techniques exploiting the wave nature of X-ray (and gamma-ray)  radiation, so excluding, for example, the use of screens with zone-plate-like patterns singly as coded masks  \cite{\mertz}  or in pairs to produce Moir\'e fringes \cite{\wilson,\desai}. Within wave optics, systems  based on multilayer optics or  on crystal diffraction are not addressed (the latter are reviewed elsewhere in this series \cite{\frontera}).

%
\section{History} 
\label{sec:history}
\begin{figure}
   \begin{center}
   \begin{tabular}{c}
 \includegraphics[trim = 0mm 0mm 0mm 0mm, clip, width=9cm]{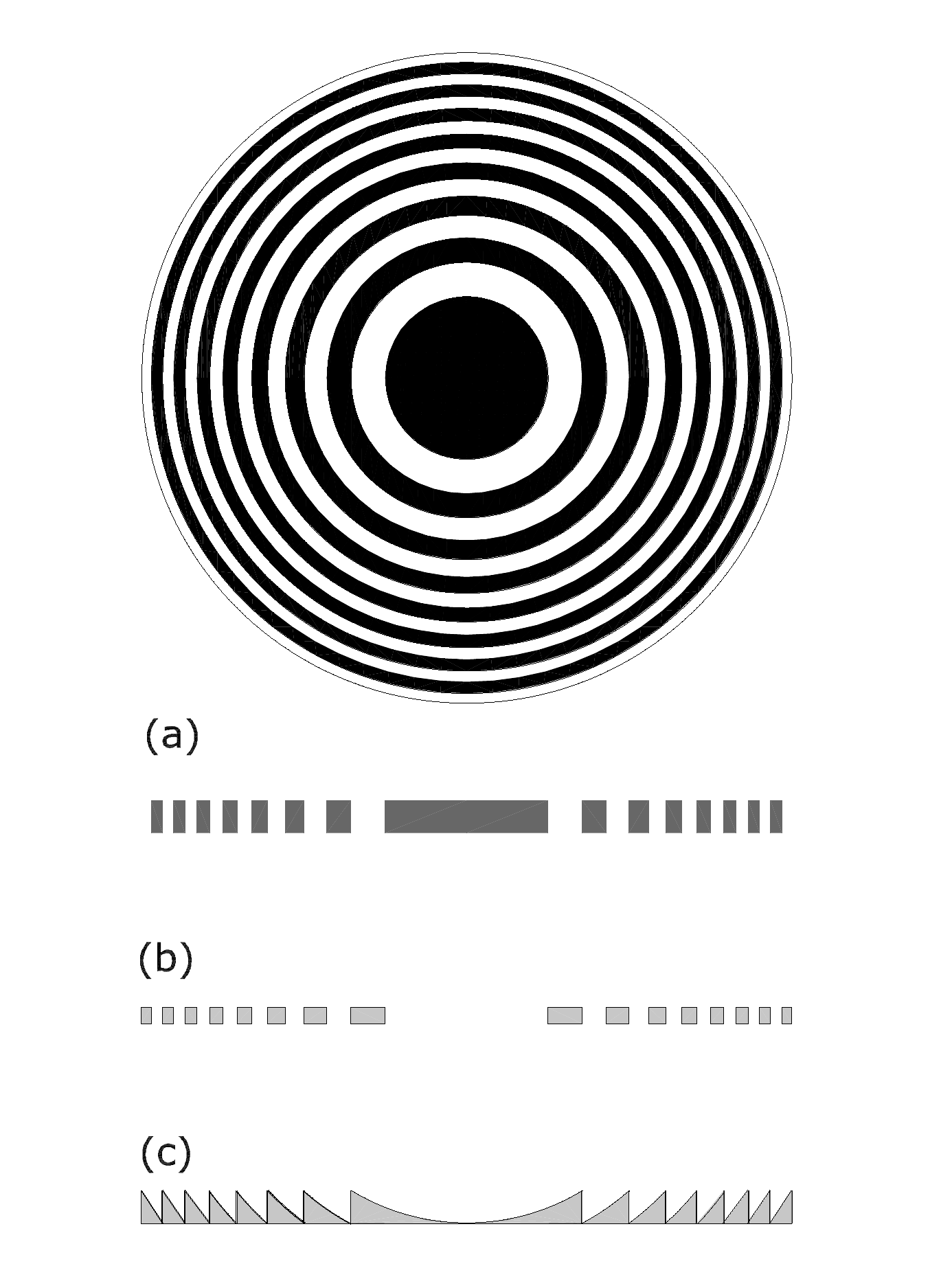}\\
   \end{tabular}
   \end{center}
   \caption[example] 
   { \label{fig:zp_pzp_pfl} 
Three basic forms of diffractive optics for X-ray telescopes. (a) A zone-plate (ZP), with opaque and transparent regions, (b) A phase-zone-plate (PZP), in which the shaded zones transmit radiation with a phase shift of $\pi$, (c) A phase-Fresnel-lens (PFL), in which the thickness is everywhere such that the phase is shifted by the optimum angle. The profile in (c) is drawn for a converging lens assuming the refractive index is less than unity.}
   \end{figure}

The basic diffractive optics imaging element can be considered to be the zone plate (ZP) in which an aperture is divided into transparent and opaque regions according to whether radiation passing through them would arrive at some selected focal point with a phase such as to interfere constructively  or destructively. The resulting pattern is shown in Fig. \ref{fig:zp_pzp_pfl}a. The first zone plate was made by Lord Rayleigh in 1871 though  this work was never published (see \cite{woodbook, \kirz}) and it was Soret \cite{\soret} who first described them in print in 1875. Already in 1888 Rayleigh\cite{\rayleigh} realized  that problems  of high  background  due  to  undiffracted  light  and    low  efficiency   could   be   overcome   by using phase-reversal zone plates (PZPs)  in which     the   opaque   regions  are  replaced  by  ones   whose thickness is chosen to introduce a phase-shift of $\pi$   (Fig. \ref{fig:zp_pzp_pfl}b).  Wood  demonstrated  the  operation  of  PZPs  ten  years  later  \cite{wood_1898}.  Miyamoto \cite{\miyamoto} further extended the concept, introducing the term phase Fresnel lens (PFL below) for an optic  in which the phase shift is at each radius the same as that for an ideal conventional lens (Fig. \ref{fig:zp_pzp_pfl}c).

The possibility of using ZPs  for X-ray imaging seems first to have been seriously considered in the 1950s by Myers \cite{\myers} and by Baez \cite{\baez}, who were concerned with X-ray microscopy.
In 1974 Kirz \cite{\kirz} pointed out that the relative transparency of materials in the X-ray band and the fact that X-ray  refractive indices differ slightly from unity  would allow  PZPs to be constructed even for high energy photons, so obtaining much higher efficiency. 
 
Remarkably, as early as the 1960s simple zone plates were used for  solar (soft) X-ray astronomy. Elwert \cite{\elwert} obtained the agreement of Friedmann to replace the pinholes in two of the pinhole cameras on a 1966 NRL solar-viewing sounding rocket flight with small zone plates designed to operate in lines at  51\AA\  (0.246 keV) and  34\AA\  (0.367 keV).
Although the attitude control malfunctioned, blurred images were obtained. Over the  next few years the technique was used, in particular by the T\"ubingen and Utrecht groups, for solar imagery from sounding rockets at energies up to 0.8 keV 
 \citep[e.g.][]{\brauninger, \burger}. One of the disadvantages of zone plates proved to be the halo due to zero-order diffraction surrounding the focal point of a simple zone plate. Ways were found of alleviating this problem by using only the zones within an annular region \cite{\dijkstra, \elwertfeitzinger}. 

At much  the same time Wolter I  grazing incidence telescopes were becoming available for soft X-ray solar imaging -- the first was flown on a sounding rocket in 1965 \cite{\giaconni} and two were used  on the Apollo Telescope mount on Skylab \cite{\vaiana}. With the size of instrumentation that was  feasible at that epoch, the grazing incidence technology proved superior. For cosmic observations as well as solar, it has become the imaging technique of choice except where the need for very wide fields of view or operation at high energies precludes its use, in which case non-focussing devices such as coded-mask or Compton telescopes are used. 

As a result of  the success of grazing angle  reflective optics,  diffractive X-ray optics for astronomical applications tended to be forgotten.
In a 1974 PhD thesis   Niemann  \cite{\niemann} did discuss the possible use of diffractive optics for extra-solar astronomy and in 1996  Dewey \etal\ \cite{\dewey}  proposed a mission concept in which patched blazed diffraction gratings based on the technology developed for the AXAF mission (now Chandra) would approximate a PZP. But it is only  comparatively recently that there has  been a revival of interest in the possibility of using diffractive optics for X-ray and gamma-ray astronomy in particular circumstances where it may offer unique advantages. Several authors  \citep[e.g.][]{\skinnera, \skinnerb, \gorensteina,  \gorensteinb, \braiga, \braigb} have pointed out that the angular resolution potentially available with diffractive X-ray telescopes exceeds by many orders of magnitude the practical limits of reflective optics.

Meanwhile there have been major advances in diffractive X-ray optics for non-astronomical applications, driven in particular by the availability of synchrotron sources and interest in X-ray microscopy with the best possible spatial resolution. For reasons discussed below, most of the effort has been towards lenses with structures on an extremely small scale, even if the diameter is also small.  For astronomical telescopes on the other hand fineness of structure would be of secondary importance, but large apertures are essential.  

%
%

\section{Theory}
\label{sec:theory} 

 \subsection{Basic parameters}
 \label{subsec:params}
 A zone plate such as illustrated in Figure \ref{fig:zp_pzp_pfl}a can be conveniently characterized by the outside diameter, $d$,  and the pitch, $p_{min}$ of one cycle of the opaque/transparent pattern at the periphery where it is smallest\footnote{The number of cycles is assumed to be large so that a local characteristic period can be defined.}. The focal length for radiation of wavelength $\lambda$ is then given by   
\begin{equation}
f = \frac{p_{min}d}{2\lambda} ,
\end{equation}
a relationship that  applies equally to PZPs and PFLs. In terms of photon energy $E$ and physical units this becomes
 \begin{equation}
f = 403.3   \left( \frac{p_{min}}{1\: \mu\textrm{m}} \right)   \left( \frac{d}{1\: \textrm{m}}   \right)   \left(    \frac {E}{1\: \textrm{keV}} \right)\: \textrm{m} ,
\label{eqn:basic}
\end{equation}
making clear that focal lengths of systems of interest for X-ray astronomy are likely to be long.
 
Diffractive X-ray optics has so far been employed almost exclusively for microscopy and for other applications for which {\it spatial} resolution is all-important, but for telescopes, it is the {\it angular} resolution that counts. For ideal PFLs one can use the usual rule that the diffraction limited angular resolution  expressed as half-power-diameter (HPD) is 
\begin{eqnarray}
  \label{eqn:diff_lim1}
  \Delta \theta_d&=&1.03\lambda/d =   263  \left(    \frac {E}{1\: \textrm{keV}} \right)^{-1}
  \left(    \frac {d}{1\: \textrm{m}} \right)^{-1}
  \: \textrm{micro-arcseconds (} \mu''\textrm{)}  \\
    \label{eqn:diff_lim}
  &=& 0.501 \left( \frac{p_{min}}{f}\right),
\end{eqnarray}
(using the Rayleigh critereon, the numerical factor 1.03 would be the familiar 1.22).
The same expression can be in practice be used for ZPs and PZPs with a large number of cycles (Stigliani et al. \cite{\stigliani} present an exact solution).   
Assuming a simple single lens system, the  corresponding dimension in the image plane is then 
\begin{equation}
\Delta x = f\;\Delta\theta_d = 0.501\; p_{min}. 
\label{eqn:dx}
\end{equation}
This is also approximately the spatial resolution of a microscope with a diffractive lens, explaining why the main drive in X-ray  diffractive optics technology has so far been towards reducing $p_{min}$. ZPs with zones of less than 15 nm, corresponding to $p_{min}<30$ nm have been reported \cite{\chao}. If the angular resolution of a  telescope is to be limited only by diffraction, then it is important that the distance $\Delta x$ be larger than the spatial resolution of the detector. Equation \ref{eqn:dx} then implies that $p_{min}$ should {\it not} be too small.  Current state of the art  detectors have pixel sizes from 5--10 \micron\ (CCDs at X-ray energies) to a fraction of a millimeter (pixelized  CZT or Ge detectors for hard X-rays and gamma-rays). Although the centroid of the released charge can in some circumstances be localized to better than one pixel,  the range of the electron which receives energy from the incoming photon sets a limit on the spatial resolution that can be obtained. So $p_{min}$ for diffraction-limited  X-ray telescopes is likely to be at least $\sim$10s to 100s of microns.
 %
 %
 \subsection{Lens profile}
 \label{subsec:profile}
 A simple zone plate is inefficient because it is only 50\% transmitting and because much of the radiation is not diffracted into the primary focus (order $n=1$) but is undiffracted  ($n=0$) or diffracted into secondary focii ($n<0, n>1$).  Table \ref{table:zp_eff} shows that even if it is perfect  the efficiency of a ZP is 
 only $\sim$10\%.
 \begin{table}[htdp]
\caption{Comparison of the efficiency of ideal zone-plates (ZPs),  phase-zone-plates (PZPs), and phase Fresnel lenses (PFLs)}
\begin{center}
\begin{tabular}{|l|c|c|c|c|}
\hline
               &    Efficiency                 & Lost              & Undiffracted    &      Secondary               \\
               &    (primary focus)        &   to                &                           &               focii                   \\
               &          $n=1$                 &   absorption  &         $n=0$      &     $n=-1,\; \pm3,\; \pm5\; ...$       \\
\hline
ZP         &     $1/\pi^2 = $10.1\%  & 50\%            &           25\%       &             14.9\%              \\
PZP      &     $4/\pi^2 = $40.5\%  &   0                 &                0         &          59.5\%                 \\
PFL      &   100\%                          &    0                 &               0         &             0                         \\
\hline 
\end{tabular}
\end{center}
\label{table:zp_eff}
\end{table}

 For a ZP the depth of the profile is simply dictated by the requirement that the material be adequately opaque;  the only upper limit is that imposed by  manufacturing considerations. High density, high atomic number, materials are generally to be  preferred. 
 Taking Tungsten as an example and assuming 45\% open area to provide for support structures, 90\% opacity implies 130 g m$^{-2}$ at 10 keV or 2.9 kg m$^{-2}$ at 100 keV.
 
 For PZPs and PFLs on the other hand, the thickness of the profile and the transparency of the material are  
  critical. The nominal thickness of a PZP is that which produces a phase shift of $\pi$. The complex refractive index of the material may be written 
 \begin{equation}
\mu  = 1-\frac{r_e\lambda^2}{2\pi}n_0(f_1+if_2)= 1-\delta -i\beta,
\end{equation}
where $r_e$ is the classical electron radius, $n_0$ is the number of atoms per unit volume, and $f=f_1+if_2$ is the atomic scattering factor.  The required thickness is then $\frac{1}{2}$\ttwopi, where \ttwopi\  is $\lambda/\delta$. Well above the energy of any absorption edges $f_1$ approaches the atomic number, so $\delta$ is proportional to $\lambda^{2}$ and \ttwopi\ to $\lambda^{-1}$, or to $E$.
   \begin{figure}
   \begin{center}
   \begin{tabular}{c}
 \includegraphics[trim = 7mm 7mm 30mm 10mm, clip, width=13cm]{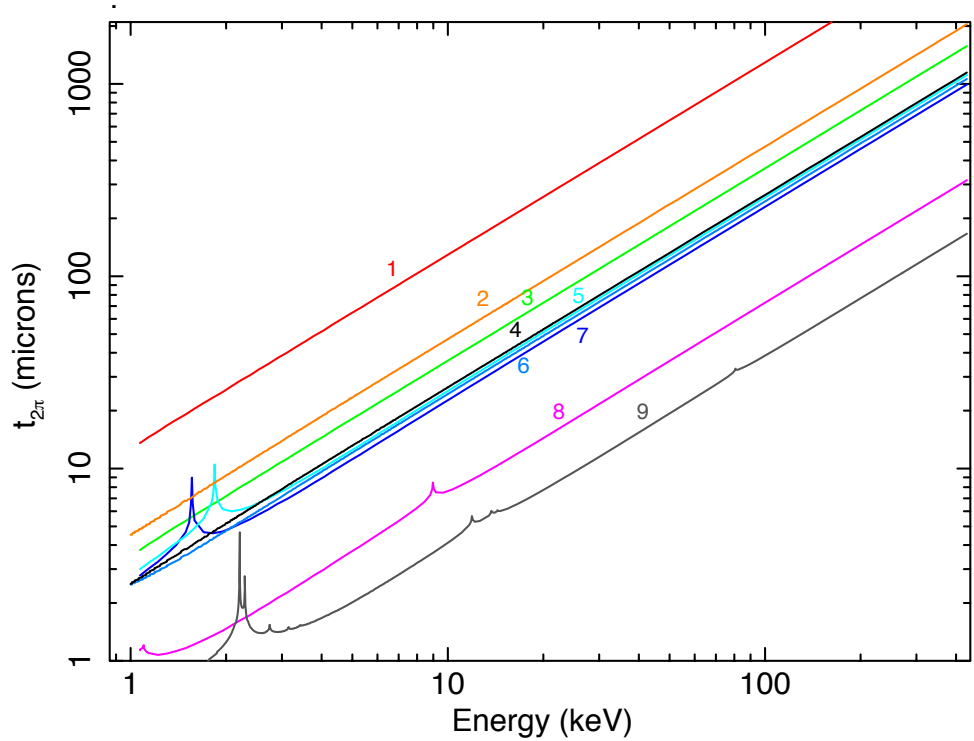}\\
   \end{tabular}
   \end{center}
   \caption[example] 
   { \label{fig:t2pi} 
The thickness, \ttwopi, necessary to shift the phase of X-rays by $2\pi$ for  some example  materials. From top to bottom  Li (1, red) Li,  Polycarbonate ( 2, orange), Be (3, green), B$_4$C (4, light blue), Si (5, cyan), LiF (6, light blue), Al (7, dark blue),  Cu (8, violet), Au (9, grey).  }
   \end{figure} 
   \begin{figure}
   \begin{center}
   \begin{tabular}{c}
 \includegraphics[trim = 7mm 7mm 30mm 10mm, clip, width=13cm]{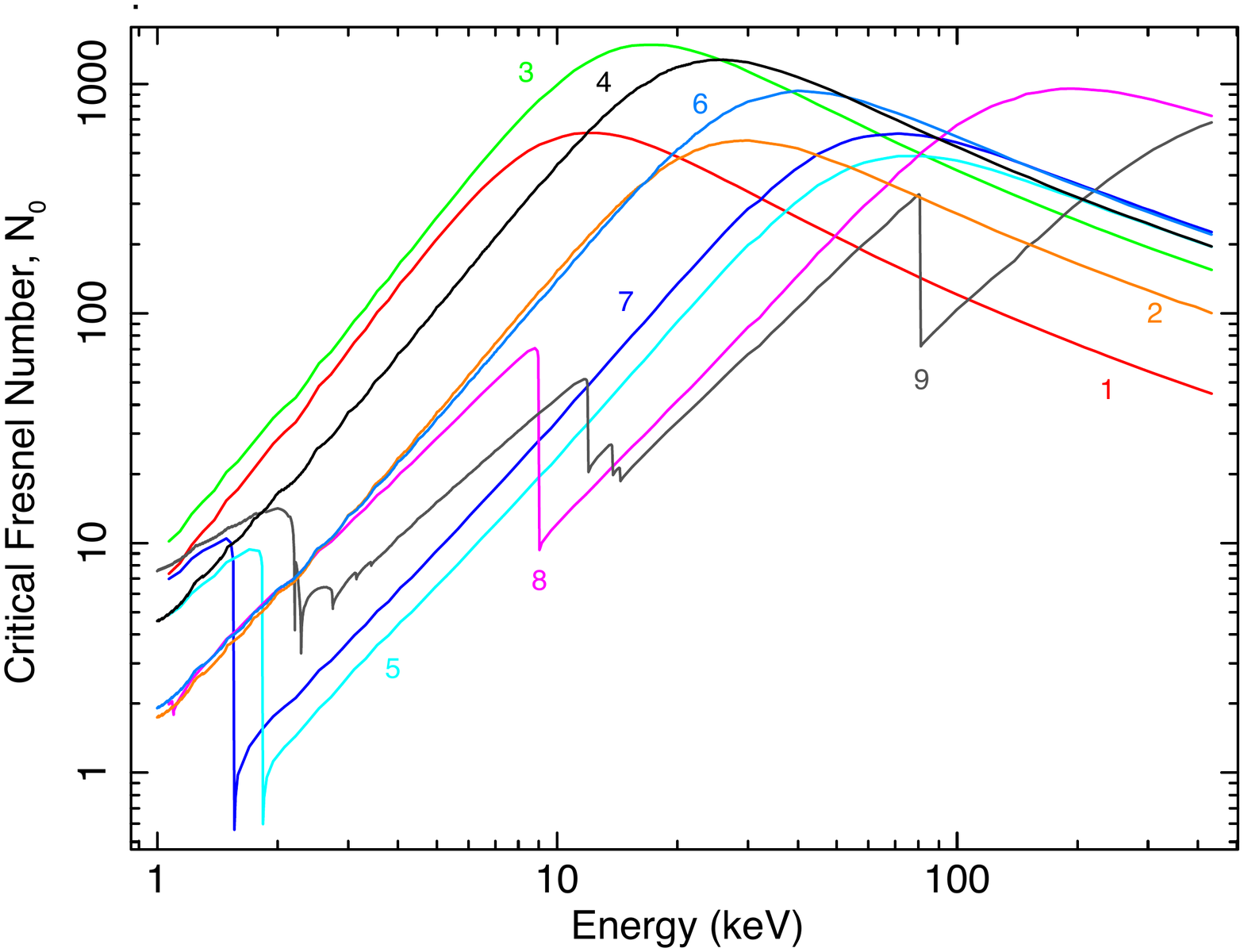}\\
   \end{tabular}
   \end{center}
   \caption[example] 
   { \label{fig:n0} 
The critical Fresnel number, $N_0$, for  some example  materials.  Numbers, colors and materials as in Fig. \ref{fig:t2pi}.}
   \end{figure} 

The practicability of X-ray PZPs relies on the fact that although the wavelengths $\lambda$ are extremely short, $\delta$ is also very small. Consequently \ttwopi\ is in a range, from microns to millimeters, where fabrication is practicable and where the absorption losses in traversing the required thickness of material are not too serious. 
 Fig. \ref{fig:t2pi} shows some values of  \ttwopi\ as a function of photon  energy for some materials of interest.  
 
 In the case of  PFLs, the surface of each ring is ideally part of a paraboloid\footnote{ This is evident if the lens is regarded as a thick refractive lens with the thickness reduced by multiples of \ttwopi.}. Usually the maximum height is \ttwopi\, though \tfourpi\, \tsixpi\ ... lenses can be made, with a correspondingly reduced number of rings.  At large radii the small sections of paraboloids are almost straight and the cross-section is close to a triangular sawtooth. With some fabrication techniques it is convenient to use a stepped approximation to the ideal profile  \citep[e.g.][]{\difabrizio}. High efficiency can be obtained with a relatively small number of levels (Table \ref{table:multilevel}). Fabrication `errors' in the form of rounding of corners will actually tend to {\it improve} efficiency.
 
 In contrast to ZPs, because of their lower absorption (see \S \ref{subsec:abs}) low atomic numbers materials will generally be preferred for the fabrication of PZPs and PFLs \footnote{The low density that tends to be associated with low atomic number can however be something of a disadvantage as it implies larger \ttwopi and so higher aspect ratio for the profile.}. In fact for a given photon energy the areal density of the active part of such devices is relatively independent of the material chosen. Typical values are only 30 g m$^{-2}$ at 10 keV or 300 g m$^{-2}$ at 100 keV. The additional mass of a substrate and/or support structure must be taken into account, but diffractive X-ray optics are intrinsically {\it very} light weight.
\begin{table}[htdp]
\caption{The primary focus efficiency, in the absence of absorption, of a stepped approximation to a PFL \citep[see][]{dammann}. }
\begin{center}
\begin{tabular}{|r|c|c|c|c|c|c|}
\hline
Levels:    &    2   (PZP)   &          3       &          4        &          8        &          16        &          $n$     \\ 
\hline
Efficiency: &   40.5\%    &     68.4\%   &     81.1\%   &       95.0\%    &    98.7\%   &    $\textrm{sinc}^2(\pi/n)$\\
\hline
\end{tabular}
\end{center}
\label{table:multilevel}
\end{table}
%
%
\subsection{Field of view}

	Young \cite{\young} has discussed the off-axis  aberrations of ZPs and the same conclusions can be applied to PZPs and in practice to PFLs. The expressions that he derives imply that in circumstances of interest for astronomy the most important Seidel aberration is coma, which only becomes important at off-axis angles greater than $4\lambda f^2/d^3$.  In other terms, this implies that the number of diffraction-limited resolution elements across the coma-free field of view is $\sim 8 (f/d)^2$. With the very large focal ratios that seem to be inevitable for diffractive X-ray telescopes, aberrations are entirely negligible over a  field of view far larger than any practicable detector.

%
 %

\subsection{Chromatic aberration and other limits to angular resolution}

\begin{figure}[htbp]
   \begin{center}
   \begin{tabular}{c}
 \includegraphics[trim = 0mm 75mm 0mm 20mm, clip, width=10cm]{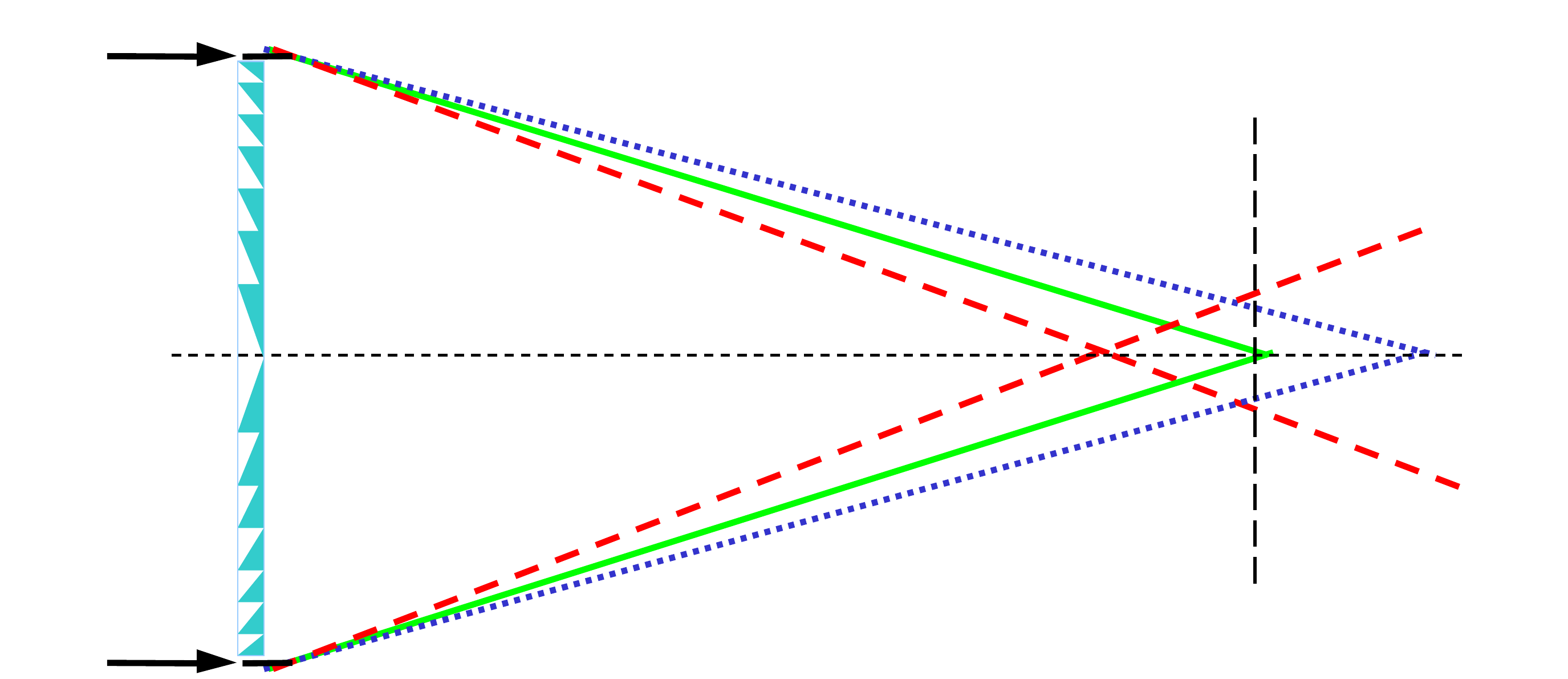}\\
   \end{tabular}
   \end{center}
   \caption[example] 
   { \label{fig:chrom} 
The effect of chromatic aberration. As the focal length of a diffractive lens is proportional to photon energy, ignoring diffraction at energies differing by $\delta E$ from the nominal energy $E$ radiation converges in a cone and intercepts the detector plane in a disc of size given by Equation \ref{eqn:theta_c}. }
\end{figure}

As well as the limit imposed by diffraction, given in Equation \ref{eqn:diff_lim}, two other considerations are important for the angular resolution. The most important is the limit imposed by chromatic aberration. It is apparent from Equation \ref{eqn:basic} that the focal length of a diffractive lens varies in proportional to the photon energy. At an energy $E'$ away from the nominal energy $E$, radiation will converge in a cone towards a displaced focus (Figure \ref{fig:chrom}). In the absence of diffraction, purely geometric considerations imply that this would  lead to a focal spot  corresponding to an angular size

\begin{equation}
\left| \frac{E'-E}{E}\right| \left( \frac{d}{f}\right).
\label{eqn:theta_c}
\end{equation}

Unless the spectrum of the radiation being imaged is a very narrow  line with negligible continuum emission, blurring due to photons with energy other than 
the nominal energy has to be taken into account. Fortunately modern X-ray detectors of interest for the low flux levels that are encountered in astronomy are photon counting and energy resolving. Thus photons of energy outside a defined bandwidth can be disregarded when analyzing the data. Consequently in the most common case of a broad line and/or continuum spectrum it is the energy resolution of the detector that has to be used for $\Delta E$ in Equation \ref{eqn:theta_c} and that dictates the degree of chromatic aberration. 

At X-ray energies the most widely used imaging detectors are CCDs, that typically have an energy resolution of $\sim$150 eV at 6 keV, corresponding to $\Delta E/E$ = 2.5\% or active pixel sensors with similar capability \cite{bautz2009}.  In the 100$-$1000 keV region Germanium detectors   can achieve   $\Delta E/E \sim$0.25--1\%, though the (fractional) resolution is not as good at lower energies.  Position sensitive Germanium detectors    
are now becoming available (see for example  \cite{\wunderer}, where the possibility of their use in a `Compton' configuration to reduce background by selecting only those events that are consistent with photons that may have passed through a lens is also discussed). Pixelated CZT arrays are approaching the performance of Germanium detectors  and do not need cooling (Li et al. \cite{\liq}  report 0.61--1.64\% at 662 keV for a variety of single pixel detectors and small arrays).
Microcalorimeter detectors are reaching energy resolutions of 2.5$-$5 eV at 6 keV \cite{\stahle, \kelley}, corresponding to $\Delta E/E\sim$0.05\%, but this performance is currently only achieved in single detector elements or small arrays. 
Braig and Predehl \cite{\braigb} have even suggested that a Bragg crystal monochromator might be used in the focal plane of a diffractive telescope and that $\Delta E/E\sim$0.01\% might be achieved in this way.

As discussed in \S\ref{subsec:params}, the spatial resolution of the detector can be important as well. The net resolution is thus in general a combination of three components
\begin{description}
\item[Diffraction: ]  { Rewriting the expression in Equation \ref{eqn:diff_lim} in terms of energy for consistency 
\begin{equation}
\Delta\theta_d=  \frac{1.03}{d}\frac{hc}{E}
\end{equation}
}
\item[Chromatic aberration: ]  Equation \ref{eqn:theta_c} allows the {\it maximum} extent of the spread due to chromatic aberration, and in the absence of effects due to diffraction, to be written 
\begin{equation}
\Delta\hat\theta_c= \frac{1}{2} \frac{d}{f}\frac{\Delta E}{E}. 
\end{equation}
\item[Detector resolution: ]  Characterizing this by a pixel size, one has 
\begin{equation}
 \Delta\theta_p =  \frac{\Delta x}{f}
 \end{equation}
\end{description}

\noindent Diffraction limited performance will only be obtainable if the second and third components are negligible compared to the first. 
The energy bandwidth  could be estimated by  comparing $\Delta \theta_d$ and $\Delta \hat\theta_c$,  but as these are different measures of the widths  of very different PSF profiles a  better  approach is  to consider the range of energies for which the wavefront error at the edge of the lens remains below some value. It is usual to place the comparatively strict limit of  $\lambda/4$  corresponding  to a Strehl ratio (the ratio between the peak brightness and that for an ideal, diffraction-limited system) of $8/\pi^2\sim80$\%. Yang \etal\ \cite{\yang} use a limit twice as large and other criteria lead to even wider bandwidths.  For $\lambda/4$ it is found \cite{\young} that
\begin{equation}
\frac{\Delta E}{E}= 4\frac{f}{d^2}\frac{hc}{E} = \frac{1}{N_F},
\label{eqn:pfl_bandpass}
\end{equation}
where $N_F$ is the number of Fresnel zone in the PFL (twice the number of periods for a \ttwopi\  lens). 
  
 %
 %

 \subsection{Absorption}
 \label{subsec:abs}
  The efficiencies discussed so far ignore  absorption. In evaluating such effects a key parameter is the critical  Fresnel number defined by Yang \cite{\yang}
  \begin{equation}
N_0=\frac{\delta}{2\pi\beta}=\frac{2}{t_{2\pi}\mu_{abs}},
\end{equation}
where $\mu_{abs}=4\pi\beta/\lambda$ is the linear absorption coefficient. $N_0$ measures the relative importance of refraction and absorption. It can be thought of as the number of Fresnel zones in a refractive lens whose maximum thickness  is equal to  the absorption length. Example values are shown in Fig. \ref{fig:n0}.

\begin{figure}[htbp]
   \begin{center}
   \begin{tabular}{c}
 \includegraphics[trim = 0mm 10mm 0mm 25mm, clip, width=9cm]{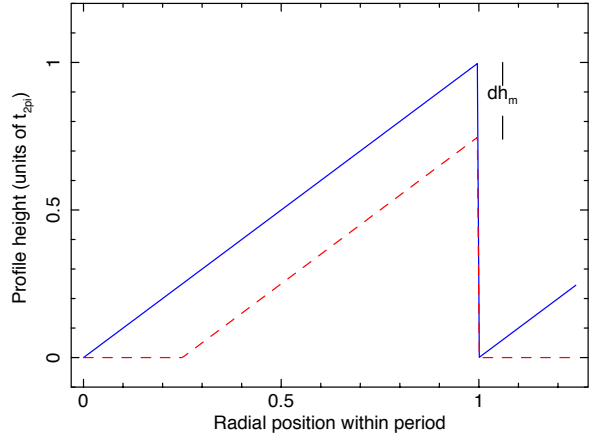}\\
   \end{tabular}
   \end{center}
   \caption[example] 
   { \label{fig:pfl_modif}    If significant absorption is taken into account the efficiency of a PFL can be improved by modifying the profile because improved throughput can be obtained at the expense of some phase error. The shape of the profile of a PFL that gives maximum efficiency for diffraction into the primary focus is shown by the dashed line.  }
\end{figure}

\begin{figure}[htbp]
   \begin{center}
   \begin{tabular}{c}
   \includegraphics[trim = 0mm 10mm 0mm 25mm, clip, width=9cm]{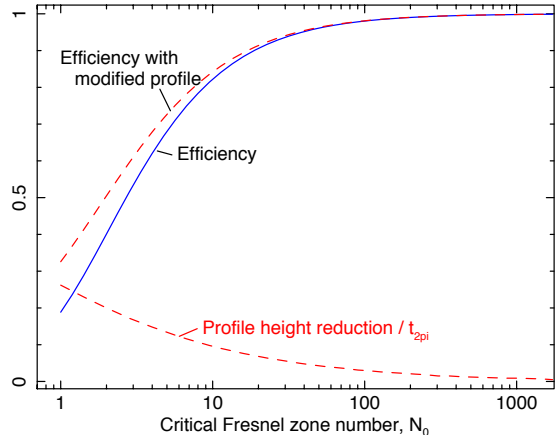}\\
   \end{tabular}
   \end{center}
   \caption[example] 
   { \label{fig:eff_v_n0}  The efficiency of a PFL with significant absorption as a function of the critical Fresnel number,  $N_0$, of the material from which it is made. Also shown are the fractional amount, $dh_m$,  by which the nominal profile height of \ttwopi\ should be reduced   as  in Fig. \ref{fig:pfl_modif} to optimize the performance, and the resulting efficiency. }
\end{figure}

Kirz \cite{\kirz} has shown how in the presence of significant absorption the optimum thickness of a PZP is somewhat lower than  $t_{2\pi}/2$. For $\beta/\delta=1$ the (primary focus) efficiency is maximized if the thickness is reduced to 0.73 of this value. Similarly it can be shown that the efficiency of a PFL made from a material in which the absorption is important is maximized if the profile is modified as shown in Fig. \ref{fig:pfl_modif}. N\"ohammer et al. \cite{nohammer} discuss this issue in more detail.  Efficiencies with and without this adjustment are shown as a function of $N_0$ in Fig. \ref{fig:eff_v_n0}. The improvement is not large, but the modified profile is actually likely to be lighter and easier to fabricate. Note that the optimum form would be different if the objective were to minimize either the undiffracted (zero-order) flux or that in higher orders, rather than maximize the flux diffracted into the prime focus.

%
%

\section{The advantages and potential} 
  \label{sec:potential}
\subsection{Fabrication and tolerances}
\label{sec:fab}
Compared with grazing incidence optics of a comparable aperture, diffractive optics for X-ray astronomy are expected to be relatively simple to fabricate. As noted in \S\ref{sec:theory}, diffraction limited diffractive telescopes are likely to have profiles with a minimum period of microns to mm -- well within the capabilities of current micro-electro-mechanical systems (MEMS) or single point diamond turning (SPDT) technologies. Because the refractive  index of the lens is so close to unity, the tolerances on the profile are comparatively relaxed. Even for the sub-nm wavelengths involved, lens figuring to $\lambda/10$ optical precision requires only $t_{2\pi}/10$ tolerances, where \ttwopi\  is $\sim$10--1000 \micron\ (Fig.~\ref{fig:t2pi}). Radial tolerances for the same precision are no tighter than $p_{min}/10$, and will usually be of a similar order of magnitude to the vertical tolerance.

Fig. \ref{fig:n0} shows that at energies  greater than a few keV it is possible to select materials with  $N_0$  greater than a few, meaning that not only is absorption of  X-rays   in the material forming the active part of a PZP or PFL  small, but the profile can be etched or machined into a somewhat thicker  substrate (or deposited onto one) without serious absorption losses.

 Thus  the profile can be close to ideal and losses can be small so, at least at their design energy, diffractive X-ray telescope  lenses can have an effective area that is very close to their geometric area, which may easily be several square meters. 

A great advantage of optical elements used in transmission rather than reflection is their relative insusceptibility to tilt errors and out-of-plane distortions. Multiple reflection mirror  systems in which the radiation undergoes an even number of reflections, such as Wolter grazing incidence optics, are better in this respect than single reflection systems (provided the relative alignment of the mirrors does not change). However they still act as `thick' lenses and if the optic (or a part of it) is tilted by angle $\delta\psi$ the resulting  transverse displacement of rays passing through it   leads to an aberration in the image on angular scale  $\sim(t/f)\delta\psi$, where $t$ is the distance between the principal planes, a measure of the thickness of the `lens'. For Wolter optics $t$ is the axial distance between the centers of the two mirrors. For the Chandra optics $(t/f)=0.08$.       Diffractive telescope lenses are very close to  ideal `thin' lenses, having $(t/f)$ ratios of $10^{-6}-10^{-9}$ or even smaller.

\subsection{Applications of  high angular resolution diffractive telescopes}

\label{subsec:resol}

The most obvious attraction of diffractive optics for X-ray telescopes for astronomy is the potential they offer for superb angular resolution.   From Equation \ref{eqn:diff_lim1} 
it is apparent that angular resolution better  than a milli-arc-second should be readily obtainable in the X-ray band. With optics a few meters in size working with hard X-rays, sub-micro-arc-second resolution should be possible.

One of the original incentives for the recent reconsideration of diffractive optics for high energy astronomy was indeed the possibility that it offers of sub-micro-arcsecond resolution. As has been discussed in proposals to use X-ray interferometry for astronomy\cite{\white, \cash}, this is the angular resolution that would be needed to image space-time around the supermassive black holes  believed to exist at the centers of many galaxies.  Even our own Galaxy, the Milky Way, apparently  harbors a black hole, Sgr A*, with a mass of 4.3$\times 10^6$ \Msun  (where \Msun\  is the mass of the sun)\cite{\gillessen}. 

The  Schwarzschild radius of a black hole of mass $M$ is
\begin{equation}
R_s=\frac{2G M}{c^2} = 2.9 \frac{M}{M_\odot} km
\end{equation}
where $G$ is the gravitational constant and $c$ the vacuum velocity of light. The radius of the `event horizon'  is  $R_s$ in the case of a non-rotating black hole or somewhat larger for one with angular momentum.  For Sgr A* $R_s$ corresponds to an angular scale of 10 \muas.  The black holes at the centers of `active' galaxies (Seyfert galaxies, Quasars and giant radio sources) can be much more massive -- for example that in M87 may be as much as 2000 times the mass of Sgr A* \cite{\gebhardt}. Their  Schwarzschild radii will be correspondingly bigger,  so despite their much greater distances (also by a factor of 2000 in the case of M87), $R_S$ in many cases still corresponds to angular scales of 0.1--10 \muas. 

\begin{figure}[htbp]
  \begin{center}
   \begin{tabular}{c}
   \includegraphics[trim = 0mm 0mm 0mm 0mm, clip, width=13cm]{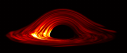}\\
   \end{tabular}
   \end{center}
   \caption[example] 
   { \label{fig:bh_sim}  Simulation of X-ray radiation from the region surrounding a black hole. A Schwarzschild black hole is assumed to be seen at an inclination angle of 80\degr. For other details see Armitage and Reynolds \cite{armitage}.  }
\end{figure}

Of course one does not expect to detect radiation from the black hole itself, but the gravitational energy released by matter as it approaches the event horizon is the origin of the extremely  high luminosity of some of these objects. Simulations have been made showing how such radiation originating near to the event horizon, even from behind the black hole itself, should appear after being bent by the gravitational field (Fig. \ref{fig:bh_sim}).  The distribution expected depends not only on the mass of the black hole but on its angular momentum and inclination angle.


\begin{figure}[htbp]
   \begin{center}
   \begin{tabular}{c}
   \includegraphics[trim = 25mm 20mm 0mm 0mm, clip, width=14cm]{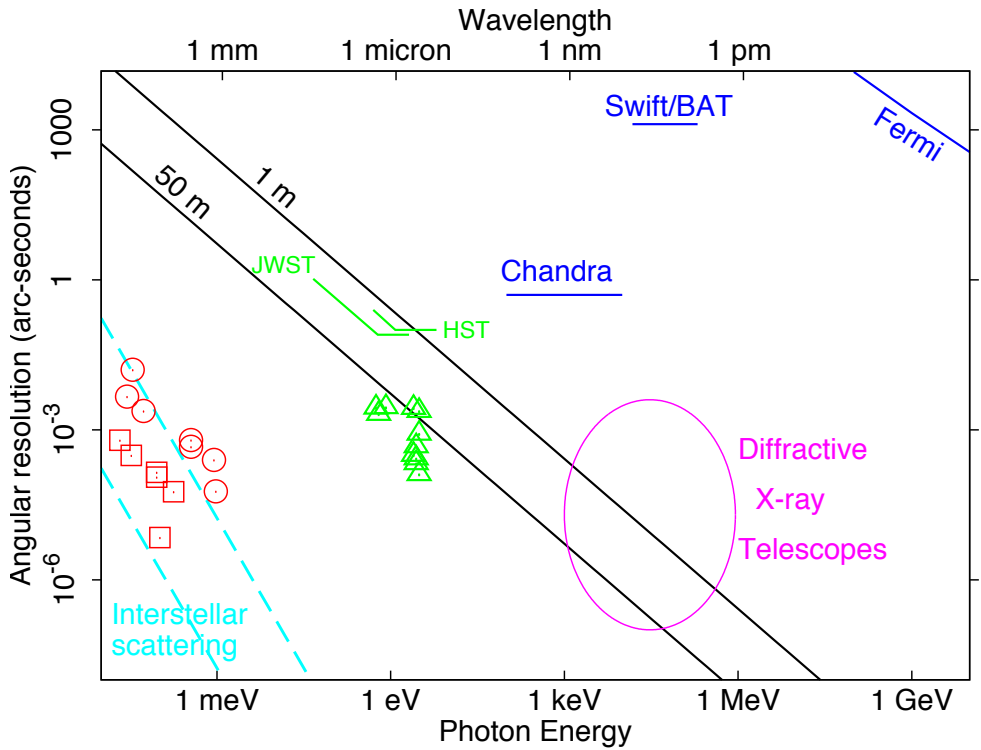}\\
   \end{tabular}
   \end{center}
   \caption[example] 
   { \label{fig:resol_comparison}  The angular resolution obtainable with different techniques and across the electromagnetic spectrum.  The approximate domain in which diffractive X-ray optics are potentially of interest is indicated  by an ellipse. Shown for comparison are (i) the Rayleigh limit for diffraction limited optics of 1 m and 50 m diameter (black lines),  (ii) the angular resolution of current X-ray instruments (blues lines)  and of the Hubble and James Webb  Space Telescopes (green lines), (iii) the best resolution of some example optical interferometer systems (green triangles) and of some typical radio VLBI measurements  (red circles),  (iv) the diffraction-limited angular resolution possible with space VLBI  (actual and near future; red squares) (v) dashed lines (cyan) roughly indicating the region in which interstellar scattering becomes dominant  at high galactic latitudes (left) and towards the galactic center (right). Note that while continuous lines refer to imaging instruments, the various symbols indicate the finest fringe spacing of interferometers which are not truly imaging.  }
\end{figure}

Fig. \ref{fig:resol_comparison} indicates the angular resolution  available to astronomers with the current state-of-the-art. 
At  present the best angular resolution  is obtained at mm wavelength by VLBI (very long baseline interferometry, see \cite{\middelberg} for a recent review). Current technologies, with transatlantic baselines and wavelengths as short as 1 mm lead to angular resolutions down to 30--40 \muas, though so far only  with a limited number of stations \cite{\krichbaum, \doeleman}, so  a characteristic  size is  measured rather than forming an image. The Russian  RadioAstron  space VLBI mission,  following in  the footsteps of HALCA/VSOP, is due to be launched  in late 2010 or early 2011 and will extend baselines to 350,000 km, but the shortest  wavelength is 13.5 mm.  Although this should allow an angular resolution of 8 \muas,  the actual  resolution will be limited by interstellar scattering except at high galactic latitude. Consequently Sgr A* cannot be observed with the highest resolution. The Japanese-led Astro-G/VSOP-2 mission will go down to 7 mm in wavelength \cite{\murata} and so be less affected by interstellar scattering (which is proportional to $\lambda^2$, as indicated by the dashed line in Fig, \ref{fig:resol_comparison}), but with the maximum baseline limited by an apogee  of only 25,000 km  the best resolution will be 38 \muas. Astro-G launch has been delayed until at least 2013 due to technical problems with the deployable dish.

A useful basis of comparison across wavebands  is the maximum baseline (or, for filled aperture instruments, the aperture diameter) in units of wavelength. Optical and infrared interferometers are pushing to higher  and higher angular resolution, though as for VLBI,  a limited number baselines provide sparse $u-v$ plane coverage and allow model fitting, but only an approximation to true imaging. 
 The 640 m and 330 m baselines of the SUSI and CHARA optical interferometers correspond to about 1.5 G$\lambda$ and 0.7 G$\lambda$ respectively, while with radio VLBI fringes have been obtained with baseline as long as 6 G$\lambda$ \cite{\krichbaumb}.  A modest  1 m diameter lens working with 6 keV X-rays would have a size of 50  G$\lambda$, and larger lenses and those working at higher energies have corresponding greater values. In addition such lenses would provide full $u-v$ plane coverage up to this scale.

It is ironical that in the X-ray and gamma-ray bands, where  the wavelengths are shortest and diffraction least limiting, the angular resolution at present obtainable
is actually inferior to that possible at longer wavelengths.  No currently planned X-ray mission will improve on, or even equal, the 0.5\arcsec\   resolution of the Chandra grazing incidence mirror. While being subject  to some constraints and limitations, discussed below, in appropriate circumstances diffractive optics offer the opportunity of improvements by up to 6 order of magnitude. X-ray imaging may then move from the present arc-second domain to  the milli-arcsecond one, where stellar surfaces can be imaged and the formation of astrophysical jets examined in detail,  and to micro-arc-seconds,  just the resolution needed for black hole imaging.

\subsection{Applications of diffractive optics for light-buckets} 

Even if one does not take advantage  of the imaging capabilities of diffractive X-ray optics, they may prove useful as flux concentrators. X-ray and gamma-ray astronomy is limited both by the small number of photons often detected and by background  in the detector, mostly due to particles. The background is typically  proportional to the detector area from which events are accepted. Catching photons requires large collecting area whereas reducing background implies that the detector should   be   as small asÊ possible. Applications of X-ray `light buckets' optimized for these purposes include high resolution spectroscopy  and fast timing measurements with moderate energy resolution such as studies of quasi-periodic oscillations, allowing the observation of certain general relativity effects such as Òframe dragging in compact black hole binaries. 

At the energies where they can be used, grazing incidence reflective optics offer a solution to the problem of how to concentrate the flux from a large collecting area onto a small detector area.  Diffractive optics provide an alternative to grazing incidence mirrors -- one that can be used even for high energies, and indeed whose performance is in many respects best at high energies.  

For a concentrator one can drop the requirement of being able to resolve with the detector a spot size corresponding to the diffraction-limited angular resolution of which the optic might be capable. Smaller $p_{min}$ may be used and so the focal length reduced.  From the same geometric considerations that lead to Equation \ref{eqn:theta_c}, the bandwidth will be 
\begin{equation}
\frac{\Delta E}{E}=2\left(\frac{d_{det}}{d}\right),
\label{eqn:bucket_bandwidth}
\end{equation}
wheras, assuming ideal efficiency for the optic and the detector,   the flux is concentrated by a factor
\begin{equation}
C= \frac{A_{eff}}{A_{det}} \approx  \left(\frac{d}{d_{det}}\right)^2 \approx 4\left(\frac{E}{\Delta E}\right)^{2}.
\label{eqn:bucket_concentration}
\end{equation}
Thus where narrow spectral lines, or groups of lines, from a compact source are to be studied, significant advantages are available.

%
%

 \section{Overcoming the difficulties}
 \label{sec:difficulties}
\subsection {Minimizing chromatic Aberration}
\label{sec:chrom_aber}

Minimizing  the effects of chromatic aberration is one of the  biggest challenges to overcome in developing diffractive   X-ray telescopes. The various measures that can be taken are discussed below.

\subsubsection{Refocusing}

\begin{figure}
   \begin{center}
   \begin{tabular}{c}
 \includegraphics[trim = 10mm 10mm 20mm 20mm, clip, width=11cm]{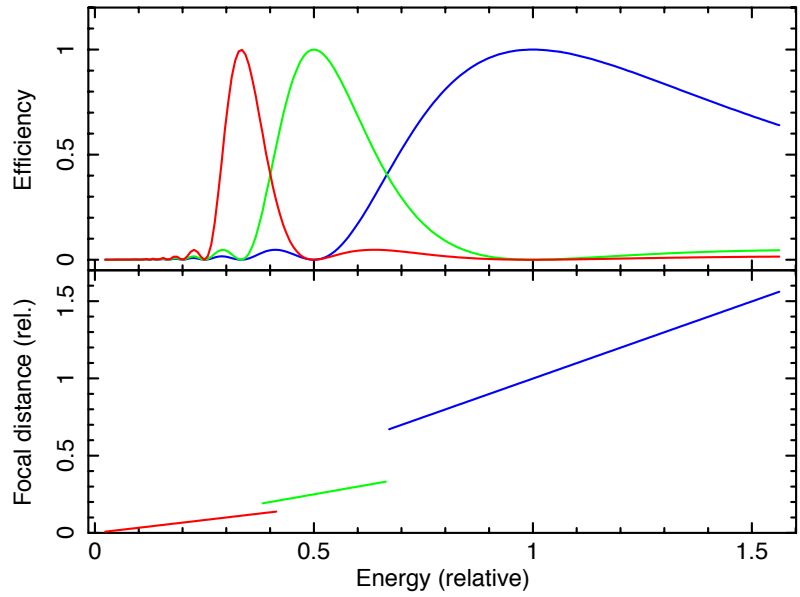}\\
   \end{tabular}
   \end{center}
   \caption[example] 
   { \label{fig:refocus} 
A PFL will work with relatively high efficiency over a broad band of energies provided that the detector plane is moved to the appropriate distance. The blue curve corresponds to the lens operating in the nominal mode. For the green and red curves the detector is assumed to be placed at the distances for the focii of the lens treated as a $t_{4\pi}$ and $t_{6\pi}$ one. Higher order responses are not  shown. Absorption effects are not taken into account in the plots. Based on \cite{\braigb}.
}
   \end{figure}

The first  thing to note is that a diffractive X-ray lenses can work well over a broad range of energies provided the detector position is adjusted for each energy.  Fig. \ref{fig:refocus} shows the effective area of a PFL as a function of energy {\it if the telescope is refocused by adjusting the  detector plane to the optimum position for each energy}.
Braig and Predehl have pointed out \cite{\braiga}  that, as indicated in the figure, well below the energy, $E_0$ for which a PFL  was designed one can take advantage of the  high efficiency  near $E_0/2$, $E_0/3$,...    because the lens  acts like a  $t_{4\pi}$, $t_{6\pi}$... one in the sense described in \S\ref{subsec:profile}.    As only one energy can be observed at a time this is not a very practical way of making broadband observations, but the approach could be useful  in studying, for example,  lines whose energy may be red-shifted to different extents  in different sources. Note that at energies where the efficiency is less than unity, the lost flux appears in focii of other orders. The resulting halo to the PSF will have low surface brightness compared with the peak, but if it is troublesome it is possible to imagine blanking off the inner  part of the lens area \cite{\braiga}.
%

\subsubsection{Improving chromatic aberration with long focal lengths}

\begin{figure}
   \begin{center}
   \begin{tabular}{c}
 \includegraphics[trim = 0mm 0mm 0mm 0mm, clip, width=10cm]{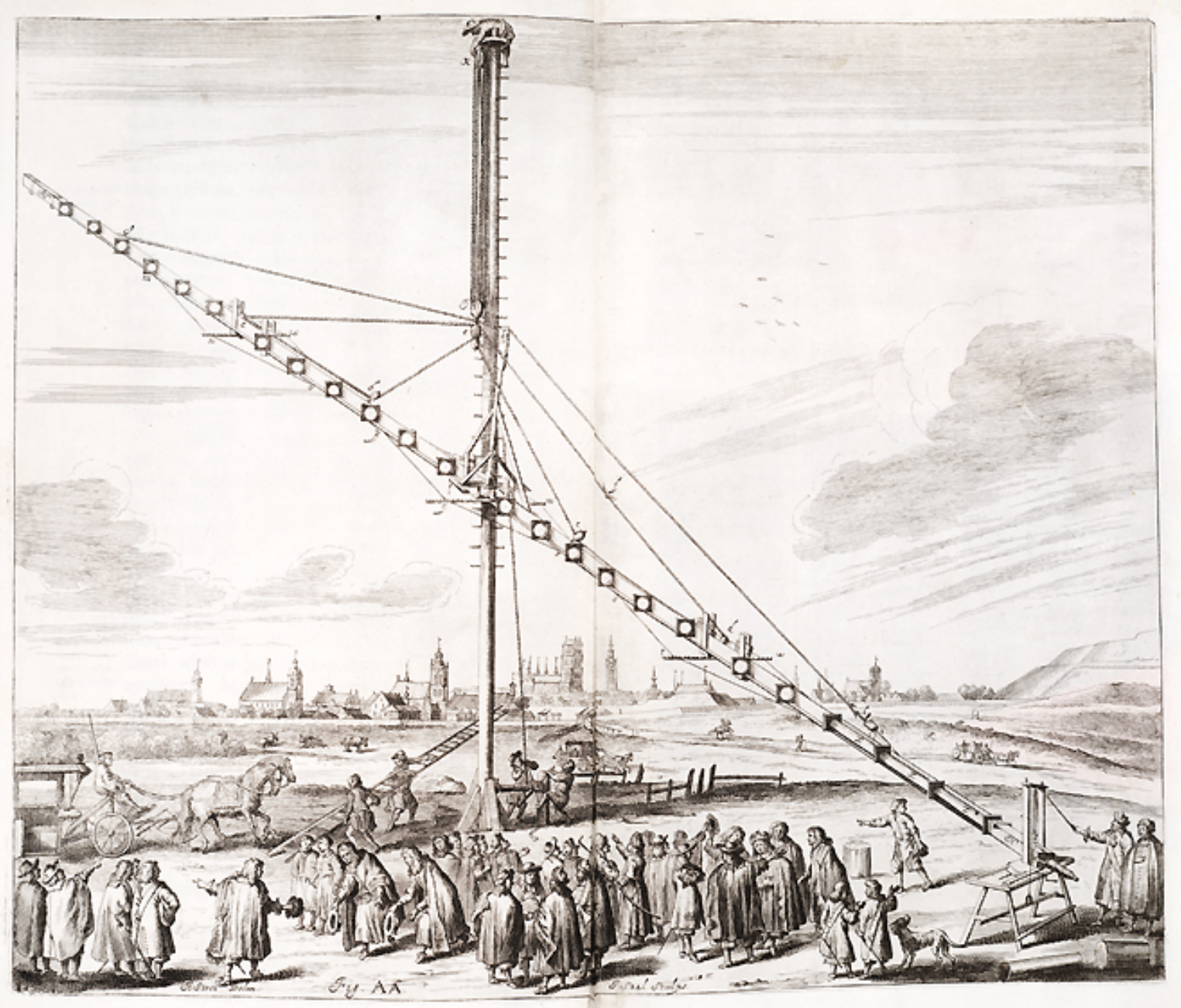}\\
   \end{tabular}
   \end{center}
   \caption[example] 
   { \label{fig:hevelius} 
It has long been known that chromatic aberration is minimized by adopting a long focal length. Before the invention of the achromatic lens Hevelius built telescopes with 60 and (shown here) 140 foot focal lengths \cite{hevelius}. }
   \end{figure}

From Equation \ref{eqn:theta_c} it is clear that the effect of chromatic aberration on angular resolution is reduced if the focal ratio is large. For a given lens size, this means that the focal length should be as long as is consistent with other constraints. This is a long-known effect;  before the invention of the (visible-light) achromatic doublet, astronomical telescopes were built with very long focal lengths in order to minimize chromatic aberration.  Fig.  \ref{fig:hevelius} shows an extreme example.  Long focal lengths lead to a limited  field of view (\S \ref{subsec:fov}) and tend to have severe implications for the logistics of maintaining the telescope in space (\S \ref{subsec:formation_flying}), but the bandwidth increases in proportion to $f$.

\subsubsection{Segmenting lenses for multi-energy operation}	

Given that PFLs of  large surface area are not too difficult to make but wide bandwidth is hard to achieve, it has been pointed out on a number of occasions \citep[e.g.][]{\skinnerb, \gorensteinb, \braiga}   that the surface of a diffractive lens can be divided into regions tuned to different energies, thus providing information over a wider bandpass. The division may be into a few zones  or into many and an aperture may be divided  radially or azimuthally, or both. The `sub-telescopes' so formed may be either concentric, or parallel but offset.

An important consideration is the interference  fringes that are produced when radiation of the same energy arrives in the same part of the detector after  passing through the different regions of the diffractive optic, including those designed for quite different energies.  In some of the work of Braig \& Predehl \cite{\braigb, \braigc, \braigd} this issue is avoided by assuming, explicitly or implicitly,  that the lens is made from many small segments that are subject to random position deviations such that their outputs add incoherently  rather than coherently. In this case the resolution is of course that associated with the characteristic size of a segment (e.g.  a few centimeters) rather than the lens as a whole ( e.g. a few meters).

\subsubsection{Diffractive-diffractive correction}

In some circumstances  the dispersion of a diffractive optical element can be corrected using that of a second diffractive optical element. However, as shown formally by Bennett  \cite{\bennett}  no optical system consisting of only two diffractive lenses can form   a real image, free from longitudinal dispersion,  from a real object. Buralli and Rogers (1989) generalized this result to any number of diffractive elements.  Michette \cite{\michette}   has described some schemes that get around this constraint, but they do so only by allowing diffraction into multiple orders with a consequent loss in efficiency and they only provide correction at two disparate energies. Other optical systems that do correct one dispersive element with another  either involve virtual images (or  objects), or  depend on reflective (or refractive) relay optics. The  space-based visible light diffractive imager proposed by Koechlin \etal\  \cite{\koechlin}) is an example of the use of reflective relay optics.

X-ray telescopes that depend on both diffractive and reflective optics risk suffering the disadvantages of both,    so correction schemes that depend on reflective relay optics do not seem very attractive. On the other hand, as will be noted below (\S\ref{subsec:focal_length}), there are perhaps other reasons for considering incorporating  a  reflective component, so perhaps a feasible diffractive-reflective-diffractive design may eventually evolve.

\subsubsection{Diffractive-refractive correction}
\label{subsubsec:diffdiff}

Given the constraints that apply to diffractive-diffractive correction for telescopes forming real images, the possibility of correction using refractive optics has been  widely considered. The constraints that preclude achromatic correction of one diffractive element with another  arise because the dispersion of the two elements  would be the same.  Correction of a diffractive lens with a glass one is used in the visible part of the spectrum where glasses have a dispersion that differs from the $E^{-1}$ dependence of the power of diffractive elements  \citep[e.g.][]{\stonegeorge}.  The same principle can be applied to X-rays -- purely refractive lenses can be made \cite{\lengeler}  and are widely used in X-ray microscopy and beam manipulation, often stacked to provide adequate power. 
Fig. \ref{fig:t2pi}  shows that for most materials over most of the X-ray band \ttwopi\  is proportional to $E$,  implying  that the power of a refractive lens, which depends on  $\delta$, is proportional to $E^{-2}$.

\begin{figure}
   \begin{center}
   \begin{tabular}{c}
 \includegraphics[trim = 0mm 0mm 0mm 0mm, clip, width=7cm]{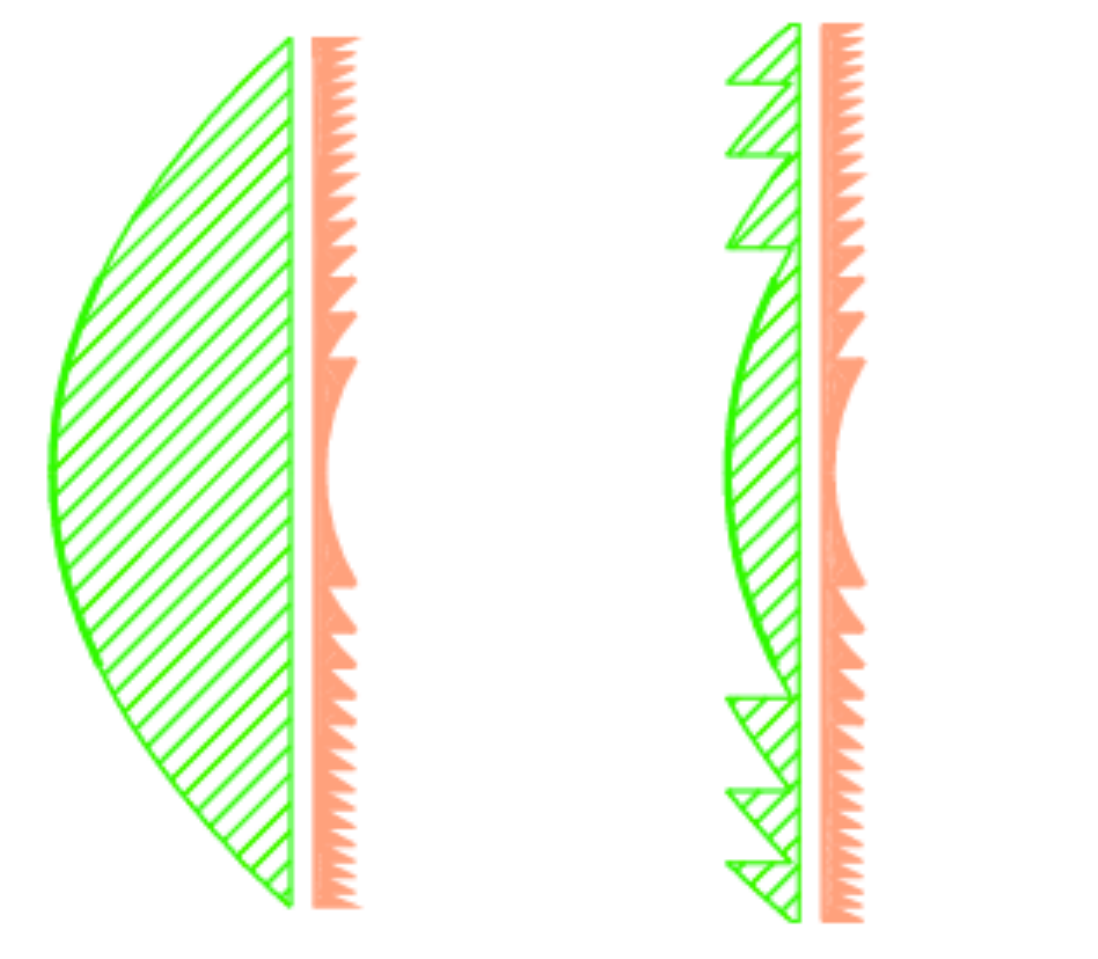}\\
   \end{tabular}
   \end{center}
   \caption[example] 
   { \label{fig:achrom} 
Left: the use of a refractive X-ray lens to chromatically correct a diffractive one. Right: a more practical configuration in which the refractive component is stepped to reduce absorption.}
   \end{figure}

Figs. \ref{fig:achrom}, shows how in principle first order correction of longitudinal chromatic aberration of a PFL (or ZP, PZP) is possible with a diffractive/refractive X-ray doublet \cite{\skinnerb, \gorensteina, \skinnerao}.   Because the lens remains a `thin' one, there is almost no lateral color aberration. In the common situation where  \ttwopi\  is proportional to $E$,  the focal length of the refractive component should be $-2f_d$, where $f_d$ is that of the diffractive one,  and the combined focal length is $2f_d$.  In this case the number of Fresnel zones in the refractive component is  half the number in the PFL.
In the absence of absorption the bandpass is increased from $\Delta E /E = 1/N_ F$ (Equation \ref{eqn:pfl_bandpass}) to $\Delta E/E = 1 / \sqrt{N_F}$  \citep[e.g.][]{\braige} 
 (here $N_F$ is the number of Fresnel zones in a single lens having the focal length of the combination).  

It has already been noted (\S\ref{subsec:abs})  that absorption becomes important in a refractive lens if $N_z$ exceeds the critical Fresnel number $N_0$ for the material. As in practice  $N_0$ is usually a few tens up to a few hundred  (Fig.\ref{fig:n0}), this sets a limit on the  size of  diffractive lens that can be corrected this way.  Wang  \etal\ \cite{\wang} have suggested using the rapid  energy dependence of $\delta$  just above absorption edges to reduce the thickness of the refractive component needed. This approach only works at very specific energies, but it could  make possible achromatic diffractive-refractive doublets  several millimeters in diameter for microscopy and microlithography. 

\begin{figure}
   \begin{center}
   \begin{tabular}{c}
 \includegraphics[trim = 0mm 0mm 0mm 0mm, clip, height=13cm]{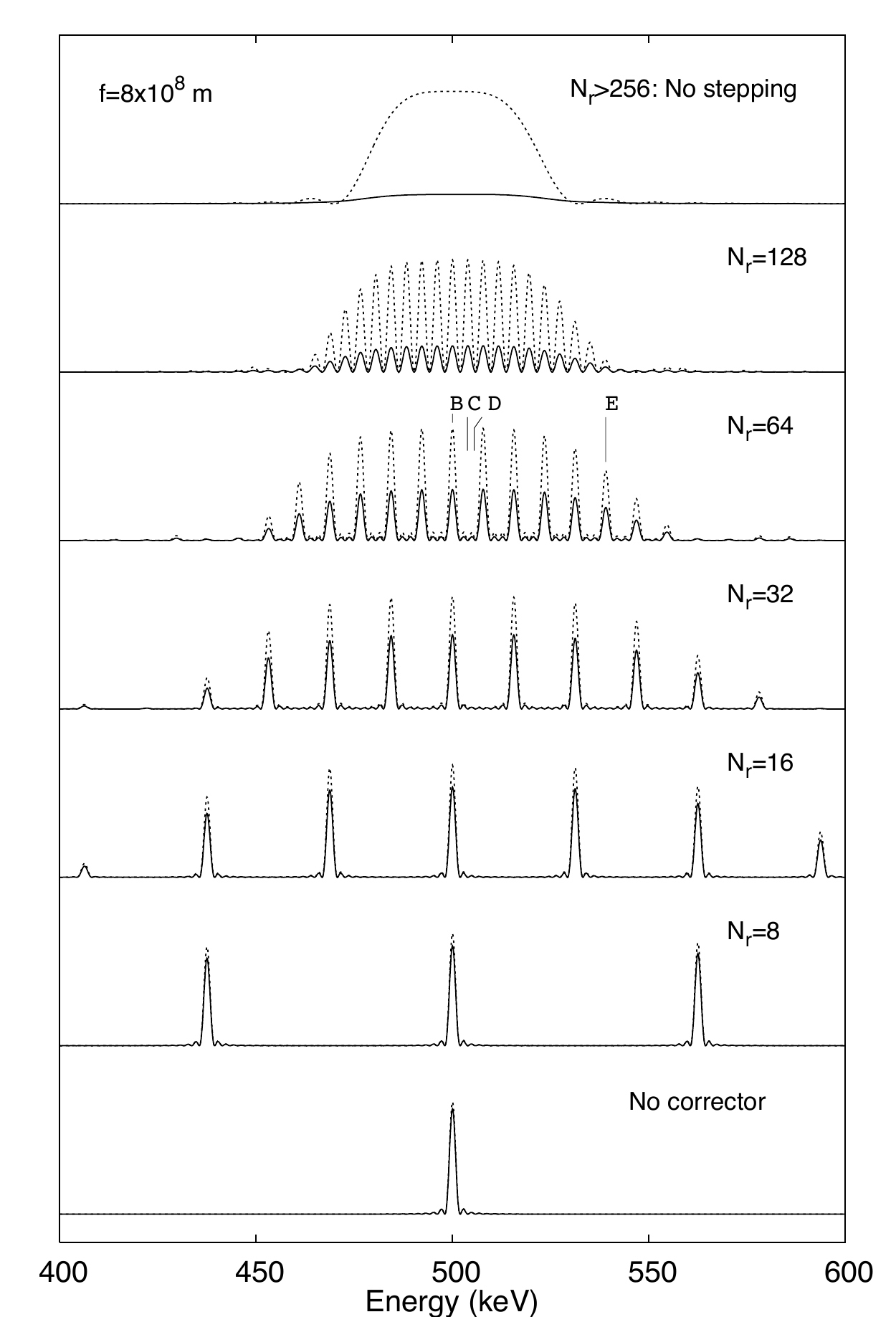}\\
   \end{tabular}
   \end{center}
   \caption[example] 
   { \label{fig:pfl_psf} 
Simulated response of a refractive/diffractive achromatic doublet with different degrees of stepping of the refractive component. 
$N_r$ corresponds to $m$ in the text. The dotted lines indicate the response if there were no absorption. As the step size is made smaller, the density of  coverage within the bandpass becomes worse but the effects of absorption become less serious and the band covered becomes wider.  Details in \cite{\skinnerao}, from which this figure is taken.}
   \end{figure}

A PFL can be considered as a refractive lens in which the thickness is reduced  $Modulo(t_{2\pi})$. Thus it is natural to consider reducing the absorption in a refractive correcting lens by stepping it back, not  $Modulo(t_{2\pi})$  which simply would make it another PFL, but  $Modulo(m\;t_{2\pi})$  for some large integer $m$ as in Fig. \ref{fig:achrom}.  \ttwopi\ varies with energy, so the value at some particular energy, $E_0$, must be chosen. Coherence is then maintained across the steps at this energy and at any other for which $m'\; [t_{2\pi}(E)] = m\; [t_{2\pi}(E_0)]$ for some other integer $m'$. This occurs at a comb of energies. In the regime where \ttwopi\ is proportional to $E$, they occur at intervals such that $\Delta E/E \sim 1/m$.  Detailed analyses of the response of PFLs with  stepped achromatic correctors have been published \cite{\skinnerao, \braigd}. Examples are shown in Fig.\ref{fig:pfl_psf}. As the step size is made smaller and the number of zones $N_Z$ within the refractive component increases, the density of  coverage within the bandpass decreases as $N_Z^{-1}$ and the band covered becomes wider (improving approximately as  $N_Z^{1/2}$) and  the effects of absorption become less serious.

\begin{figure}
   \begin{center}
   \begin{tabular}{c}
 \includegraphics[trim = 0mm 0mm 0mm 0mm, clip, width=13cm]{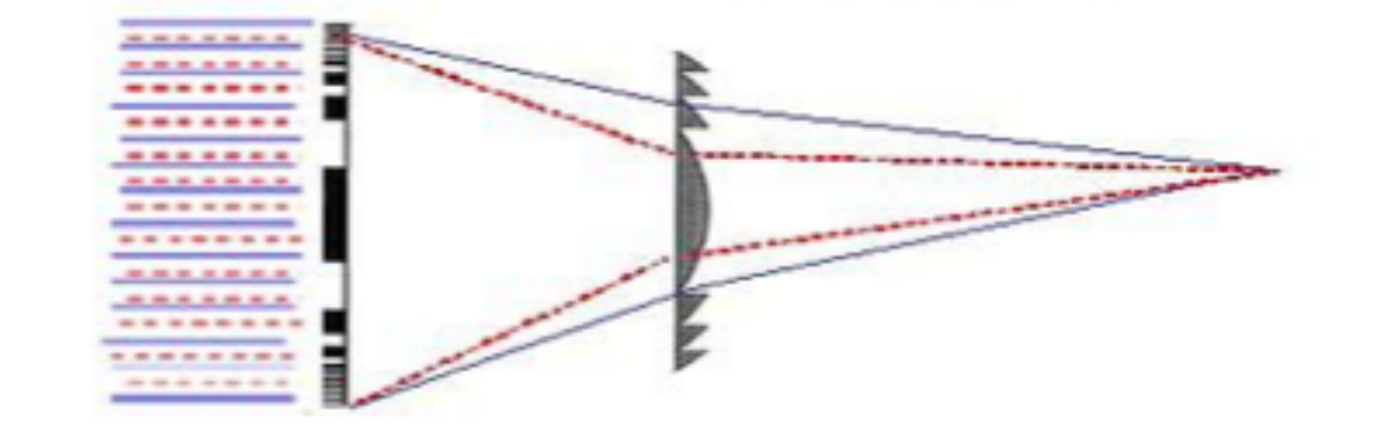}\\
   \end{tabular}
   \end{center}
   \caption[example] 
   { \label{fig:apochrom} 
A variation on the refractive/diffractive achromatic doublet shown in Fig. \ref{fig:achrom} in which the components are separated, allowing second order correction of longitudinal color (Figure from \cite{\gorensteind}). A stepped  version of the refractive component is shown.}
   \end{figure}

If one is willing to consider configurations involving the alignment  of not just two widely spaced components, but three, then the configuration indicated in  Fig. \ref{fig:apochrom} offers an even wider bandpass (see \cite{\gorensteinb}\  and references therein; more details are provided in \cite{\skinnerao}).  With the extra degree of freedom introduced by the separation between the two lenses it is possible to set to zero not just $dz/dE$ ($z$ being the axial position of the image), but also $d^2z/dE^2)$. The bandwidth is increased as indicated in Table \ref{table:achroms}. The fact that the image is somewhat magnified compared with a single lens system of the same overall length could be advantageous.  However as the magnification is energy dependent, care must be taken that lateral chromatic aberration does not limit the useful field of view (this will not be a problem if the detector has adequate energy resolution to allow post-facto scale-factor correction).

%
 \begin{table}[htdp]
\caption{Achromatic correction of PFLs. Parameters are given for systems with the same overall length, $f$, at the nominal design energy, $E_0$, and for the case when \ttwopi\ is proportional to energy, $E$. In the table  $\Delta E$ is specified relative to $E_0$ and $d$ is the diameter of the refractive component.  }
\begin{center}
\begin{tabular}{|l|c|c|c|}
\hline
                                                                     &    Simple PFL                 & Doublet             &  Separated pair            \\
\hline
 & & & \\
Diffractive  focal length        &     $f$                               &          $\frac{f}{2}$           &       $\frac{f}{3} $                  \\
 & & & \\
Separation                                 &                 -                      &                 0                              &   $\frac{f}{9}$                               \\
 & & & \\
Refractive  focal length        &           -                               &            $-f$           &        $-\frac{8}{27} f$               \\        
 & & & \\              
Image plane distance            &     $f(1+\Delta E)$                           &    $f \left(1+\Delta E^2 +... \right)$    &     $f \left(1+1.35\Delta E^3 +... \right)$                                  \\
 & & & \\
Image scale                            &     $f$                               &          $f$              &       $\frac{4}{3}f \left(1-\frac{1}{2}\Delta E +... \right)$                                  \\
 & & & \\
Diffraction-limited                               &     
                                   \multirow{2}{*}{    $  4\left( \frac{f}{d^2}\frac{hc}{E}  \right)  $   }               &      
                                   \multirow{2}{*}{      $   4\left(\frac{f}{d^2} \frac{h c}{E} \right)^{1/2} $ }   &     
                                   \multirow{2}{*}{      $  \left( \frac{16}{3} \frac{f}{d^2} \frac{h c}{E} \right)^{1/3} $   }                    \\
bandwidth (fractional)& & & \\
\hline 
\end{tabular}
\end{center}
\label{table:achroms}
\end{table}

\subsubsection{Axicons, Axilenses and other wideband variants}

Variations of the ZP have been proposed in which the shape of the PSF is modified to (slightly) improve the angular resolution \citep[e.g.][]{\simpsonmichette} or  in the case of `photon sieves'  to improve the resolution available with a given minimum feature size in the fabrication \cite{\kipp}. One such variant, the `fractal photon sieve' has been shown to have an extended spectral response  \cite{\gimenez},  achieved because some of the power from the prime focus is diverted into focii at other distances. As will be seen below, this can be accomplished in other more controlled and efficient ways.

A diffractive lens can be considered as a circular diffraction grating in which the pitch varies inversely with radius so that radiation of a particular wavelength is always diffracted towards the same focal point. Regarded in this way, a PFL is a variable pitch phase grating, blazed in such a way that all of the energy goes into the +1 order. Skinner \cite{\skinnerk, \skinnerl}\ has discussed the application to X-ray telescopes of designs in which the pitch varies radially  according to laws other than $r^{-1}$. They can be considered as  forms of radially segmented lenses in which the pitch varies smoothly rather than in steps.

 In an extreme case the pitch is constant and one has the diffractive X-ray axicon.
Interestingly, to a first approximation the point source response function of an  axicon is independent of energy. According to the Rayleigh criterion, the angular resolution is similar to that of a PFL of the same diameter working at the lowest energy, but the secondary diffraction rings are stronger so the HPD  is larger. Indeed over a wide range of radii the PSF shape is such that the enclosed power  simply increases,  approximately linearly, with radius.  In some respects the system should be regarded as an (achromatic) interferometer in which the information is contained as much in the fringes as in the central response.

\begin{figure}
   \begin{center}
   \begin{tabular}{c}
 \includegraphics[trim = 0mm 10mm 0mm 25mm, clip, height=70mm]{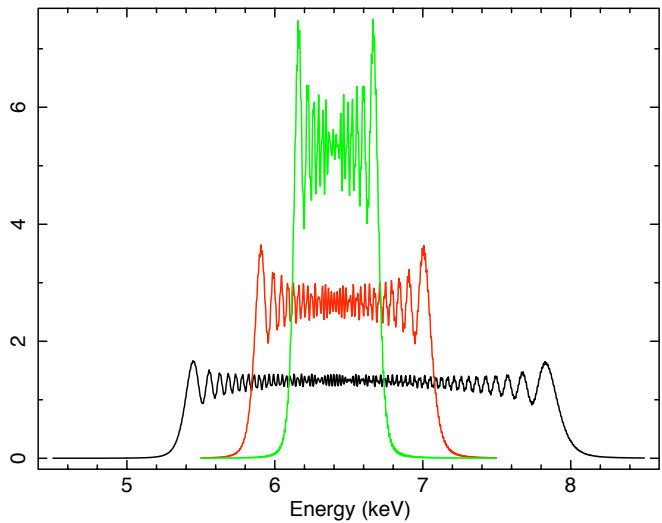}\\
   \end{tabular}
   \end{center}
   \caption[example] 
   { \label{fig:axilens_bandpass} 
 The on-axis response as a function of photon energy at a fixed distance  of 100 m for Axilenses with different parameters \cite{\skinnerk}. }
   \end{figure}

Intermediate designs are possible in which the bandpass is tailored to particular requirements based on the ideas of Sochacki \cite{\sochackia} and of Cao \& Chi \cite{\cao}. This leads to the diffractive X-ray axilens discussed in \cite{\skinnerk}. The bandwidth can be selected at will;  Fig. \ref{fig:axilens_bandpass} shows the response of some examples.  For diffraction-limited operation, the area-bandwidth product tends to be no greater than that of a narrow band PFL. As the later has the advantage that the detector noise within a narrow band is small, devices of the axicon/axilens family are most likely to find application where an extended bandpass is essential, for example in order to include the whole of a broadened line or a group of lines.

\subsubsection{Figures of merit}

Braig and Predehl  \cite{\braige} define an `Achromatic Gain' parameter $\mathcal{G}$ to measure the advantage that an achromatic system has over a simple lens. It is essentially the ratio of the effective areas, integrated over the  energy range in the two cases.

The achromatic gain  $\mathcal{G}$ may be generalized  to quantify the advantage compared to a simple PFL  of other variants as well as of achromatic combinations. As  neither the resolution nor the image scale are  necessarily the same in all cases, it is best to base the effective area on the flux within a diffraction-limited focal spot rather than peak brightness.  Such a  figure of merit was used in \cite{\skinnerl} where it was shown that axicons and axilenses  have a ` $\mathcal{G}$'  close to unity --  the extra bandwidth is obtained at the expense of effective area. The same is obviously true if a lens is segmented into areas devoted to different energies. On the other hand increasing the focal length is accompanied by an increase in bandwidth (Equation \ref{eqn:pfl_bandpass}) and of  $\mathcal{G}$. 

Care must be used in interpreting this parameter.  It measures the (square of) the improvement in signal-to-noise ratio on the assumption that the noise is dominated by photon statistics  from the source. If the observation is dominated by detector background that is not source-related, such as that due to particles, then a dependence on the square-root of the bandwidth would be more appropriate\footnote{ Provided of course that the passband of the optic is wider than the energy resolution of the detector}. In the detector background limited case, the background will also depend on the detector area over which the signal is spread \footnote{ Provided in this case that the focal spot  is larger than the detector spatial resolution}.  Furthermore, by considering only the central spot, the information carried in by the flux  high amplitude sidelobes such as occur in the PSF of axicons and axilenses, is ignored.  

\subsection {Focal length; Formation flying}

\label{subsec:formation_flying}
\label{subsec:focal_length}
 Equation \ref{eqn:basic} makes it clear that focal length, $f$,  of any practical system is likely to be long, particularly at high energies.  In addition in various  respects discussed above,  {\it very} large $f$  will often give the best performance. 
 
 First, in section  \S\ref{subsec:params} it was seen that a minimum focal length is required if spatial resolution of a detector at the prime focus is not to be a limiting factor. For example even if the detector pixels are as small as 10 \micron\, a resolution of 1 \muas\ implies $f>2000$ km. Gorenstein \cite{\gorensteinc} has suggested that this problem might be alleviated if a grazing incidence reflecting telescope were used to re-image the prime focus, perhaps with a magnification by a factor of 20.  Radiation from a particular sky pixel  would in practice illuminate a very small part of the telescope surface because the latter would be very close to the focal plane. The reflecting surface  would in effect remap positions over its aperture to pixels in a detector plane. 

A second reason that long $f$ may give the best performance is that  as noted in \S\ref{sec:chrom_aber}  chromatic aberration is minimized if $f$ is long.

Finally, based on Equation \ref{eqn:basic}, unless $f$ is large diffractive lenses of a reasonable size will have a very small period $p_{min}$ and fabrication will be more difficult and perhaps less precise.

For focal lengths of up to $\sim$100 m, it is perhaps possible to consider telescopes with a boom connecting the lens and the detector assembly. Deployable booms of up to 60 m have been flown in space \cite{\mast}; the record for the largest rigid structure in space is now held by ISS at over 100 m. 
On the other hand  even for a focal length of 50 m,  ESA studies for the proposed XEUS grazing incidence mirror X-ray mission concluded that a boom was unnecessary and that formation flying of separate spacecraft carrying the optic and the detector assemble offered a more viable solution \cite{\xeus}.

Formation flying for a long focal length telescope mission implies maintaining two spacecraft such that the line joining them has a fixed direction in inertial space. Because of gravity gradients within the solar system, as well as disturbances such as those due to differences in solar radiation pressure, a continuous thrust will be needed on at least one spacecraft.  Another consideration is that a major repointing of the telescope will require maneuvering   one of the spacecraft by a distance $\sim f$. 

Some of the issues associated with formation flying of two spacecraft for a diffractive X-ray telescope mission have been briefly discussed in the literature \cite{\skinnera, \braiga}.   Internal studies at  NASA-GSFC's Integrated Mission Design Center  of possible missions based on this technique have considered the issues more deeply. Krizmanic \etal\ \cite{\krizmanicb} has reported on one of these and updated some of the conclusions.  The missions studied  were considered ambitious, but possible. A single launcher would launch both  lens and detector spacecraft either to the vicinity of the L2 Lagrangian point   or into a `drift-away' orbit around the sun. Existing ion thrusters can provide the necessary  forces both to maintain the pointing direction against gravity gradient forces and disturbances and for re-orienting the formation. The fuel and power needed for the thrusters are acceptable. In short, no `show-stoppers' were identified.

   \begin{figure}
   \begin{center}
   \begin{tabular}{c}
 \includegraphics[trim = 0mm 0mm 0mm 0mm, clip, width=10cm]{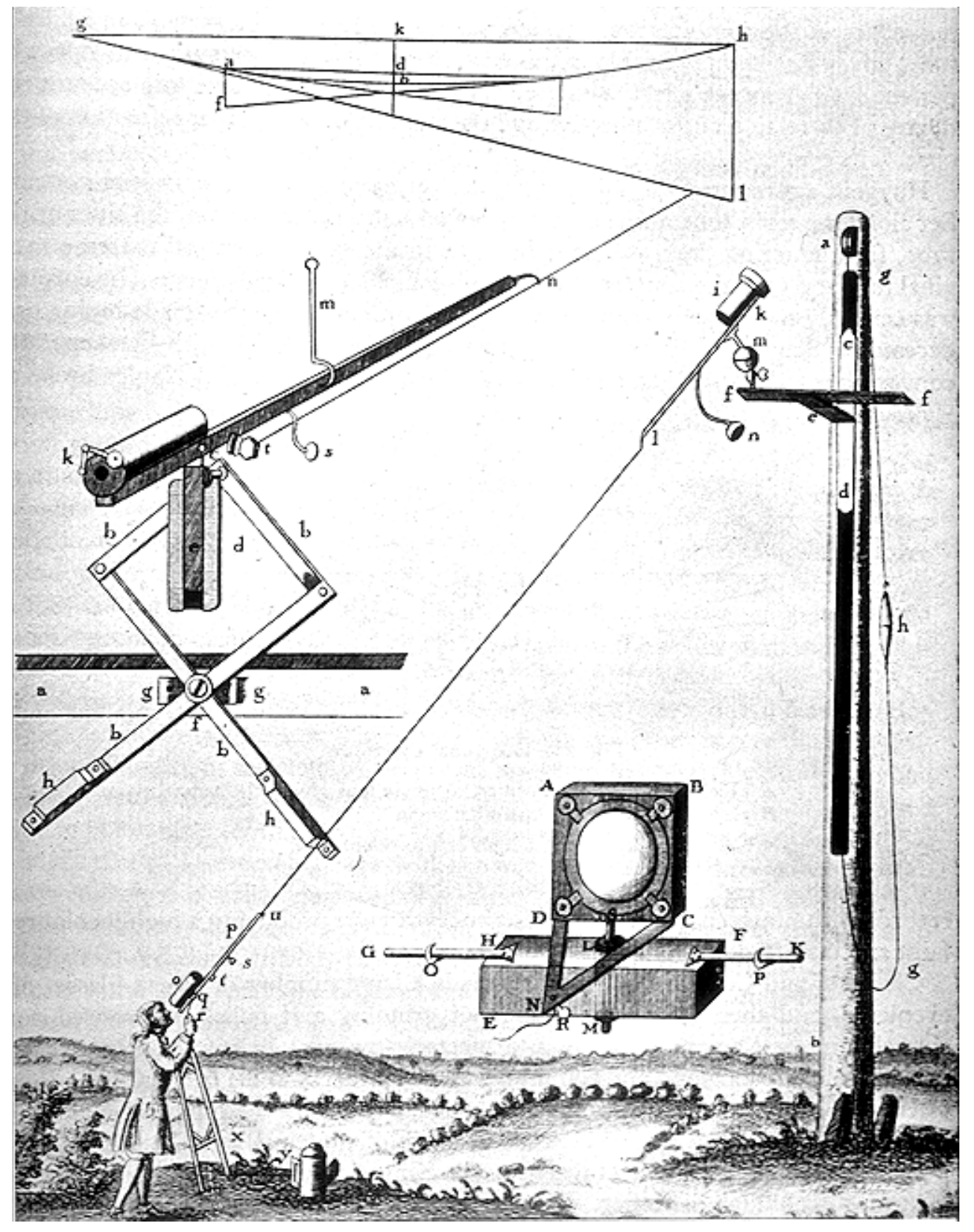}\\
   \end{tabular}
   \end{center}
   \caption[example] 
   { \label{fig:huygens_aerial} 
The concept of a telescope in which the objective of a telescope and the detector are not be rigidly connected  is not new. Christian Huygens' (1629-1687)  
used this 210 foot focal length telescope \cite{huygens_aerial}. Note that the design takes advantage of the relative immunity of a thin lens to tilt errors.  }
   \end{figure} 

\subsection{Pointing knowledge}

Although precise control of the {\it distance} between lens and detector is not needed, the knowledge of the {\it direction} in celestial space of their  vector separation is crucial. As information about every photon detected can be time-tagged, {\it control} is needed `only' to the extent necessary to ensure that the image of the region to be studied falls on the detector area (\S\ref{subsec:fov}). {\it Knowledge} of the orientation is needed, however, to a level corresponding to the angular resolution required. This amounts to a need to establish the transverse position of the detector with respect to a line passing from the source through the center of the lens, and to do so with a precision that can be from a millimeter down to a few microns. 

The problem of precise attitude determination in celestial coordinates is one common to all missions attempting to work at the milli-/micro- arcsecond level. It has already been considered in the context of the studies of MAXIM, a proposed mission to address the problem of imaging space-time around black-holes using X-ray interferometry. Gendreau \etal\ reviewed a number of approaches to the problem \cite{\gendreaua}. If a  laser  beacon is placed on the lens spacecraft and viewed against background stars, a `super-startracker'  on the detector spacecraft can in principle provide the necessary information. Obtaining measurements with the necessary precision is not out of the question -- astrometry is already in an era where milli-arc-scond accuracy  is the norm and micro-arc-second the target. 
The problem is obtaining that precision with faint stars on short enough time-scales. Fortunately various technologies discussed by Gendreau \etal\  offer  the prospect of  gyroscopic systems that would have sufficient precision to allow interpolation between absolute fixes from the stellar observations. To further alleviate the difficulties, Luquette and Sanner \cite{luquettesanner} have discussed how knowledge of the dynamics dictating the spacecraft displacements can help with determining short term changes to the alignment.

\subsection{Field of View}

The fields of view of  ultra-high angular resolution telescopes such as those possible using diffractive optics are necessarily  limited.  With a single lens configuration, or with a contact achromatic doublet, the field of view will simply be the detector size divided by the focal length.
 With long focal lengths, reasonable detector sizes imply small fields of view. Although optical systems with separated lenses (\S \ref{subsubsec:diffdiff}) or with relay mirrors (\S \ref{subsec:focal_length}) can change the image scale, this would probably be in the sense of increasing it,  so reducing the field of view.
 
 There is, anyway, a basic limitation. If for example micro-arc-second resolution were wanted over a 1\arcsec\  field of view, a detector with $>10^{12}$ pixels would be needed. The present state of the art for X-ray detectors in space is indicated by a few examples.  The EPIC camera on XMM-Newton has an array of 7 CCDs with a total of  $2.5\times 10^6$ pixels.  The imaging array of the ACIS focal plane instrument on Chandra has $4\times 10^6$   pixels in 4 CCDs. Although the e-Rosita telescopes will have a total area of CCDs about an order of magnitude greater than present generation X-ray instruments, the pixel size will be larger to match the telescopes and the number of pixels will be only just over $10^6$ \cite{\erosita}.  Instruments planned for the IXO  X-ray observatory are also in the few megapixel range, though a single monolithic device will cover 100 cm$^2$ \cite{treis}.
 
 Larger arrays of larger format  CCDs are possible. Although the requirements are a little different for visible radiation,  ESA's  Gaia astrometry mission will have an array of 106 CCDs covering about half a square meter,  with $\sim9\times10^8$ pixels \cite{\gaia}.  The Gaia pixels are rectangular $20\times 30$ \micron. Taking 20 \micron\ as a typical dimension, a similar array would provide a field of view 30 milli-arc-seconds across with 1 \muas\ pixels at a focal distance of just over 4000 km.
  
 Of course, even were a $10^{12}$ pixel detector available,  it would require an extremely strong source to produce a significant signal in even a small fraction of the pixels. Sources of interest for study at the resolution possible with diffractive X-ray optics are necessarily rather compact.
 
\label{subsec:fov}
%
%

\section{Alternatives to simple lenses} 
\label{sec:alternatives}

Equation \ref{eqn:diff_lim1} implies that  sub-micro-arcsecond angular resolution should be possible with modest sized lenses working at a few hundred keV.  There is a special interest, though, in such measurements in the X-ray band. Not only are the photons more numerous, but the emission spectra of AGN typically show strong emission lines in the region of 6.7 keV associated with highly ionized Fe. These lines carry important diagnostic information because they are shifted in energy both by gravitational redshifts  and by the Doppler effect. Ultra-high angular resolution mapping at energies in this band would be particularly valuable but would requires lenses ~ 50 m or more in diameter.

Although membrane/unfoldable Fresnel lenses of 25-100 m diameter  for visible light have been proposed \cite{\hyde} and even been demonstrated on the ground on scales of up to 5 m \cite{hydedixitSTR}, X-ray lenses of this size would not be easy to engineer and would actually have an effective area larger than needed from the point of view of photon flux.
    
A PFL  with a partially filled aperture can be envisioned. It could comprise subsections of a PFL mounted on multiple spacecraft distributed over a plane \cite{\skinnerm, \skinnerk}.
The system then becomes somewhat similar to that proposed within the studies of the proposed MAXIM (Micro-Arcsecond-X-ray IMager)  mission in which mirror assemblies on spacecraft distributed over a plane would divert radiation to form fringes on a detector situated at  a large distance \cite{\maximref}.  In both cases tight control of the distances of the spacecraft from the center of the array would be needed (to a fraction of $p_{min}$ for the partially filled PFL; similar for a MAXIM formation of the same size).
The differences are (1) that  the subsections of a  partially filled aperture PFL  would concentrate flux whereas the plane mirrors proposed for MAXIM would not and (2) mirror reflection is achromatic whereas the  PFL subsections would divert radiation  by a wavelength-dependent angle and fringes would only formed where the concentrated, deflected, beams cross. The bandwidth over which fringes appear can be improved by altering the radial dependence of the pattern pitch in the PFL in ways analogous  to X-ray axicons and axilenses. In the constant pitch (axicon) case  one has a MAXIM-like interferometer in which the beam diverters are blazed diffraction gratings  and  the borderline between imaging and interferometry becomes blurred.

%
%

\section{Conclusions; Status and prospects} 
\label{sec:conclusions}

As mentioned in the introduction, diffractive X-ray telescopes presently exist almost entirely as concepts and proposals on paper. Some work has been conducted on verifying that no problems exist and on demonstrating feasibility using scaled systems \cite{\krizmanica} fabricated by gray scale lithography \cite{morgan}. By scaling down in radius (but not in thickness) the form of a large PFL that might be used for astronomy, the focal length is reduced. Lenses a few mm in diameter and with a focal length of  $\sim$100 m can act as models of ones, say,  a few meters in diameter for which $f\sim$100  km. With this reduced focal length testing is possible using existing ground-based facilities such as the 600 m long X-ray interferometry testbed at NASA-GSFC \cite{\arzoumanian,testbed} Ironically the smaller lenses are  {\it more} difficult to make than a flight lens would be, because the period of the profile is correspondingly reduced. 

 ~ Tests and developments of small-scale PFLs, using various fabrication techniques and including achromatic combinations are continuing \cite{currentwork}. Progress to a micro-arc-second mission would probably be through an intermediate level `pathfinder mission'.  Braig \& Predehl \cite{\braigd} have suggested a concept in which a 3.5 m  lens is  divided radially into two regions  operating  respectively at 5 and 10 keV. Each  would be  made from  many small achromatic segments whose outputs are assumed to add incoherently so the angular resolution is  $\sim$1 milli-arcsecond\footnote{The present author doubts the practicability of the proposed use of Lithium for the 5~keV component due to fabrication difficulties and to oxidation.}.   
 
  A specific proposal for a pathfinder mission, MASSIM  \cite{\skinnern},  operating in the milli-arcsecond regime  has been made in the context of NASA's `Advanced Mission Concept Studies'. With a focal length of 1000 km, MASSIM would provide sub-milli-arc-second resolution  with an effective area of 2000-4000 cm$^2$ over a pseudo-continuous  5-11 keV band using  five 1-meter diameter achromatic lenses.   MASSIM would  allow important scientific objectives to be achieved, in particular allowing the imaging of the inner regions of jets where the acceleration takes place, while at the same time providing a stepping-stone to an eventual micro-arc-second mission. It requires two spacecraft , one station keeping with respect to the other another  to a fraction of a meter. A major technology driver would be the determination of direction of the actual line of sight with an accuracy corresponding to the target resolution. The proposed pointing determination method involves a state-of-the-art startracker on the detector spacecraft viewing the sky and simultaneously a beacon on the spacecraft carrying the lenses. 
 
 Another possible route forward is through  solar imagery where even small diffractive lenses, little larger than those already demonstrated and having modest focal lengths ($< $100 m), could provide images of Fe line emission $\sim$6.7 keV   from active regions with an  angular resolution many times better than any yet achieved.

In conclusion, diffractive optics offer a new family of possibilities for  telescopes. Despite drawbacks in terms of inconvenient focal lengths, and limited  field of view and bandwidth,  the potential that   they have for  flux concentration even at high energies and, in particular,   for superb angular resolution suggests that in due course they will find their place in the range of techniques available to   X-ray and gamma-ray astronomy.

 {\it Acknowledgements}
The author wishes to thank  colleagues for helpful discussions and in particular John Krizmanic for a careful reading of the manuscript and for useful suggestions.





\bibliographystyle{elsarticle-num}
\bibliography{zp_library,extra_refs}







\end{document}